\documentclass[prl,superscriptaddress,twocolumn,aps,floatfix,longbibliography,nofootinbib]{revtex4-1}
\usepackage{pslatex,graphicx,dcolumn,bm,natbib,amssymb,amsmath,color,mathtools}
\usepackage[utf8]{inputenc}
\usepackage[T1]{fontenc}
\usepackage{soul}
\usepackage{float}
\usepackage{comment}
\usepackage{ulem}
\newcommand{\be}{\begin{equation}}
\newcommand{\ee}{\end{equation}}

\newcommand{\mdm}[1]{{\color{black}#1}}

\newcommand{\beginsupplement}{%
        \setcounter{table}{0}
        \renewcommand{\thetable}{S\arabic{table}}%
        \setcounter{figure}{0}
        \renewcommand{\thefigure}{S\arabic{figure}}%
     }

\makeatletter
\def\frontmatter@thefootnote{%
 \altaffilletter@sw{\@fnsymbol}{\@fnsymbol}{\csname c@\@mpfn\endcsname}%
}%
\makeatother

\begin{document}

\title{Epithelial layer fluidization by curvature-induced unjamming} 

\author{Margherita De Marzio}
\email{Corresponding author: nhmdm@channing.harvard.edu}
\affiliation{Channing Division of Network Medicine, Brigham and Women's Hospital and Harvard Medical School, Boston, MA 02115, USA}
\affiliation{Harvard T.H. Chan School of Public Health, Boston, MA 02115, USA}
\author{Amit Das}
\affiliation{Department of Biochemical Engineering and Biotechnology, IIT Delhi, New Delhi, India}
\author{Jeffrey J. Fredberg}
\affiliation{Harvard T.H. Chan School of Public Health, Boston, MA 02115, USA}
\author{Dapeng Bi}
\email{Corresponding author: d.bi@northeastern.edu}
\affiliation{Department of Physics, Northeastern University, Boston, MA 02115, USA}

\begin{abstract}
    
    The transition of an epithelial layer from a stationary, quiescent state to a highly migratory, dynamic state is required for wound healing, development, and regeneration. This transition, known as the unjamming transition (UJT), is responsible for epithelial fluidization and collective migration. 
    Previous theoretical models have primarily focused on the UJT in flat epithelial layers, neglecting the effects of strong surface curvature characteristic of the epithelium \textit{in vivo}. In this study, we investigate the role of surface curvature on tissue plasticity and cellular migration using a vertex model embedded on a spherical surface. Our findings reveal that increasing curvature promotes the UJT by reducing the energy barriers to cellular rearrangements. Higher curvature favors cell intercalation, mobility, and self-diffusivity, resulting in epithelial structures that are malleable and migratory when small, but become more rigid and stationary as they grow. Together, these results provide a conceptual framework to better understand how cell shape, cell propulsion, and tissue geometry contribute to tissue malleability, remodeling, and stabilization.  
\end{abstract}
\maketitle

To heal a wound, remodel tissues, or invade surrounding structures, the confluent
epithelial layer transitions from a sedentary, quiescent state to a highly migratory
and dynamic one\cite{friedl2009collective,rorth2009collective}. This phenotypic
switch has been interpreted as the unjamming transition (UJT)\cite{atia2018geometric,
park2016collective,park2015unjamming,mitchel2020primary,atia2021cell,oswald2017jamming,
park2016cell,pegoraro2016problems,sadati2013collective,de2021genomic,huang2022shear,
yang2021configurational,hertaeg2024discontinuous,cai2022compressive,cai2024matrix}. In the jammed phase, cellular rearrangements are rare because cells are locked in place by
neighbors, leading to tissue rigidity. In the unjammed phase, cells move cooperatively
with neighbors, leading to tissue fluidity and malleability\cite{angelini2011glass,
nnetu2013slow,lin2021energetics,garrahan2011dynamic}.

Theoretical models have characterized the UJT in epithelial layers on flat surfaces\cite
{alert2020physical,bi2015density,bi2014energy,bi2016motility,vicsek1995novel,henkes2011active, chiang2016glass,czajkowski2019glassy,li2019mechanical}. By contrast, epithelial layers \textit
{in vivo} often reside on strongly curved surfaces, where the radius of curvature can be as small
as several cell diameters. Examples include spherical pulmonary alveoli, tubular small airways,
and ellipsoidal embryos\cite{yevick2015architecture,chen2019large,vassaux2019biophysical}. 
Experiments show that such curved geometries alter epithelial plasticity: topological defects and
out-of-plane forces impact cell packing and cytoskeletal organization\cite{glentis2022emergence, brandstatter2023curvature,yevick2015architecture,chen2019large,werner2018mesoscale,rougerie2020topographical,xi2017emergent,tomba2022epithelial,yevick2014effects,tang2022collective}. 
Curvature influences migration and cooperativity, inducing coherent rotational
motions and cell alignment\cite{glentis2022emergence,brandstatter2023curvature,xi2017emergent, tomba2022epithelial,tang2022collective}.

Previous theoretical models have studied tissue plasticity of curved epithelia in the contexts of the developing neural tube, the fly leg disc, and the mouse optical cup\cite{fletcher2014vertex,alt2017vertex,monier2015apico,inoue2016mechanical,eiraku2011self}. However, these particular geometries make it difficult to isolate the intrinsic effects of surface curvature from those induced by topological constraints. 
As such, the mechanistic role of surface curvature on cell packing, mobility, and the resulting unjamming dynamics remains poorly understood.

To fill this gap, here we embed a vertex model (VM)\cite{park2015unjamming,bi2014energy,barton_avm_2017,das2021controlled,hertaeg2024discontinuous,li2023nature} onto a spherical surface. Using this spherical VM, we investigate the UJT within physiological ranges of cell density and surface curvature. We show that as surface curvature progressively increases, energy barriers to cellular rearrangements progressively decrease.  In the presence of cellular propulsion, increasing curvature favors cell migration, larger self-diffusivity, and more frequent intercalation events.

Our study is significant in several ways. Firstly, it suggests that developing curved
structures are unjammed, fluid-like, and malleable when small but become more jammed,
solid-like, and rigid as they grow. Secondly, it identifies a novel mechanism of epithelial
layer fluidization: curvature-induced unjamming. Finally, it offers a quantitative framework to understand how cell shape, propulsion, and tissue geometry together contribute to the migratory phenotype \textit{in vivo}.

\begin{figure}[h]
\includegraphics[width=\linewidth]{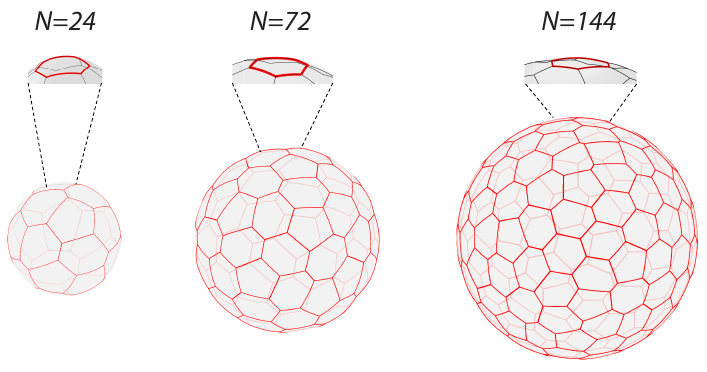}
\caption{Spherical VMs with various cell numbers $N$, corresponding to sphere radii $R=1.38, 2.39, 3.38$. Spheres with lower $N$ exhibit higher surface curvature.
}

\label{fig:1}
\end{figure}

\paragraph{The Spherical Vertex Model.}
We embed a 3D apical vertex model\cite{bi2015density,bi2014energy,alt2017vertex, fletcher2014vertex,nagai2001dynamic,farhadifar2007influence} onto a spherical
topology (Fig.\ref{fig:1}). The apical surface of cells is specified by the
vertices' locations, constrained to the spherical surface. Cell edges are defined
by geodesic curves connecting adjacent vertices. We define the system's mechanical
energy as $E=\sum_{i=1}^N{[K_A(A_i-A_0)^2+K_P(P_i-P_0)^2]}$. $A_i$ and $P_i$ are the apical area and perimeter of the \textit{i-th} cell, calculated from the geodesic polygons enclosed by the vertices\cite{fletcher2014vertex,farhadifar2007influence,staple2010mechanics,bi2015density,bi2016motility}. $A_0$ and $P_0$ are
the preferred area and perimeter, homogeneous across all $N$ cells. $K_A$ and $K_P$
are the elastic moduli for area and perimeter deformations. This energy function
accounts for cell incompressibility\cite{hufnagel2007mechanism,farhadifar2007influence},
tissue bulk and actomyosin cortex elasticities\cite{hufnagel2007mechanism,zehnder2015cell,farhadifar2007influence,staple2010mechanics}, and the balance between cell-cell
adhesion and cortical tension\cite{staple2010mechanics,farhadifar2007influence, bi2015density,li2019mechanical}. We set $A_0=\bar{A}$, with $\bar{A}$ the average
cell area. Choosing $K_P\bar{A}$ as the energy unit and $\sqrt{\bar{A}}$ as the length
unit, the energy functional reduces to:
\begin{equation}
\label{energy}
    \epsilon=\sum_{i=1}^N{[k_A(a_i-1)^2+(p_i-p_0)^2]},
\end{equation}
where $a_i=A_i/\bar{A}$ and $p_i=P_i/\sqrt{\bar{A}}$ are the rescaled cell area and
perimeter. $k_A=K_A\bar{A}/K_P$ is the rescaled area elasticity and $p_0=P_0/\sqrt{
\bar{A}}$ is the dimensionless target shape index.

We examine the spherical VM at various radii of curvature while isolating the influence of number density. We therefore choose $R=\sqrt{{N}/{4 \pi}}$  such that each cell has mean unit area. Consequently, spherical surfaces with fewer cells $N$ exhibit higher curvature (Fig.\ref{fig:1}). Different system sizes $N=36-360$ are analyzed, corresponding to radii of curvature in the range of $1.69 - 5.35$ times the mean cell size. This constitutes a physiologically relevant range experienced by cells in curved epithelia\cite{tang2022collective}.
A range of $p_0$ and $k_A$ values is studied at each $N$. For each parameter choice, we use a combination of the first-order conjugate-gradient and second-order Newton's method to find the nearest local energy minima\cite{brakke1992surface,kelley1999iterative}(SM\cite{suppl_mat}). For each equilibrium configuration, we perform T1 transitions on each edge to calculate the corresponding energy barrier.
Using the same energy minimization scheme, we simulate also a flat VM of $N=300$ cells. Finally, we examine the spherical VM in the presence of cell motility at various curvatures and fixed $p_0=3.65$(SM\cite{suppl_mat}). \mdm{This $p_0$ value ensures the single-cell dynamics to remain Brownian for all analyzed $N$ values.}

\paragraph{Curvature promotes epithelial fluidity.}We first study how substrate geometry influences tissue rigidity. In the case of flat surfaces, the vertex model predicts a rigidity transition at the critical value $p_0=p_0^*\sim3.81$\cite{bi2015density,bi2014energy,bi2016motility,li2019mechanical,yan2019}. This transition is driven by a percolation transition of the tissue's tension network, defined as the network of edges with tension $\tau>0$\cite{li2019mechanical,yan2019,petridou2021rigidity}. We examine the spatial organization of the tension network in the spherical VM(SM\cite{suppl_mat}).

\begin{figure}
\includegraphics[width=\linewidth]{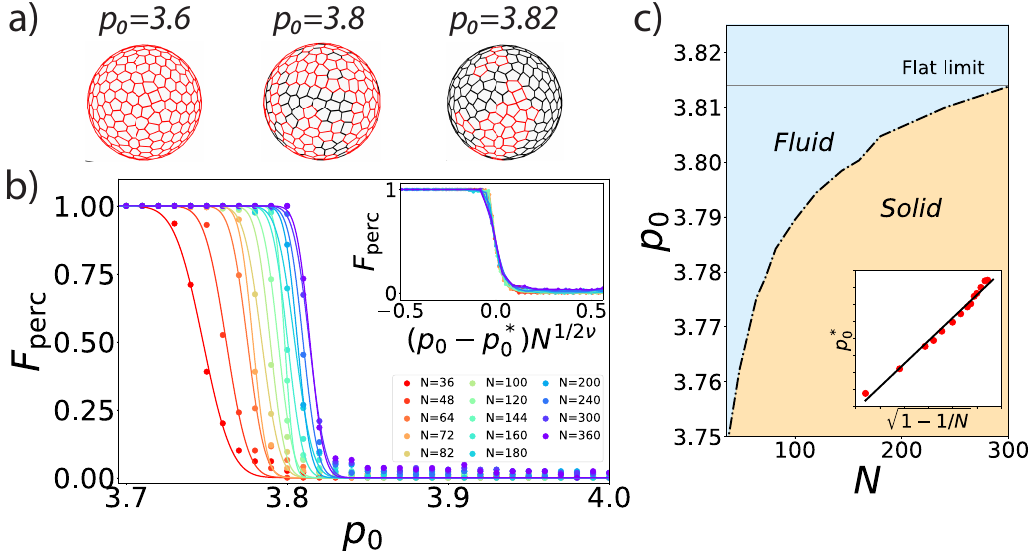}
\caption{a) The tension network at $N=300$, $k_A=1$, and various $p_0$. Red edges have finite tension. b) Probability of tension percolation $F_{\text{perc}}$ versus $p_0$ for different $N$ at $k_A=1$. $F_{\text{perc}}(p_0,N)$ is well described by $1/2\{1-\text{erf}[(p_0 - \mu_N)/(\sqrt{2}\sigma_N)]\}$ (solid lines). When $F_{\text{perc}}$ is rescaled by finite-size effects ($N^{1/2 \nu}$, $\nu=0.6$) and shifted by the percolation threshold $p_0^*(N)$, $F_{\text{perc}}$ collapses onto a single curve (inset). c) Phase diagram of the spherical VM. The percolation threshold $p_0^*(N)$ (dashed line), marking the onset of rigidity, increases with $N$.
(Inset) In agreement with our predictions, $p_0^* \propto\sqrt{1-1/N}$.
}
\label{fig:2}
\end{figure}
At fixed curvature, the tension network displays a behavior analogous to the flat surface. At lower $p_0$ (Fig.\ref{fig:2}a left panel), the network forms a system-spanning structure and the tissue is completely rigid. By increasing $p_0$, the network segregates progressively into disconnected components that are reminiscent of a fluid-like phase\cite{li2019mechanical} (central and right panels of Fig.\ref{fig:2}a). We quantify this behavior at different curvatures by calculating the probability of tension percolation $F_{\text{perc}}$ at various $p_0$ and $N$ (Fig.\ref{fig:2}b, SM\cite{suppl_mat}).  
Similar to the flat tissue, $F_{\text{perc}}$ decays with $p_0$ as an error function. $F_{\text{perc}}$ becomes sharper and more heavy-tailed by increasing $N$, as expected due to finite-size effects\cite{li2019mechanical}. Increasing $N$ also shifts $F_{\text{perc}}$ towards higher values of $p_0$, suggesting a role of curvature that goes beyond finite-size scaling. In confirmation of this role, the tension percolation threshold $p_0^*(N)$, and hence the onset of rigidity, decreases by increasing the surface curvature (Fig.\ref{fig:2}c, SM\cite{suppl_mat}). In the zero-curvature limit $N\rightarrow\infty$, $p_0^*(N)$ approaches the flat limit $p_0^*(\infty)\sim3.814$. $F_{\text{perc}}$ collapses onto a single curve after shifting by $p_0^*(N)$ and rescaling by the finite-size scaling law $N^{1/2 \nu}$, with $\nu=0.6$ (Inset of Fig.\ref{fig:2}b). Notably, the shift in the onset of rigidity can be understood within the framework of geometric compatibility. Increasing curvature in our model is analogous to applying isotropic strain: higher curvature uniformly compresses cell perimeters while keeping cell areas fixed. This analogy allows us to derive the dependence of $p_0^*$ on $N$ using recent findings on under-constrained spring networks\cite{Lee_Merkel_sm,merkel2019minimal}(SM\cite{suppl_mat}). Indeed, we can predict $p_0^*(N)$ to scale as $p_0^*(N)\propto \sqrt{1-1/N}$. Our simulations confirm this prediction, as shown in the inset of Fig.\ref{fig:2}c. These results indicate that curvature promotes tissue fluidization.

\begin{figure}
\includegraphics[width=\linewidth]{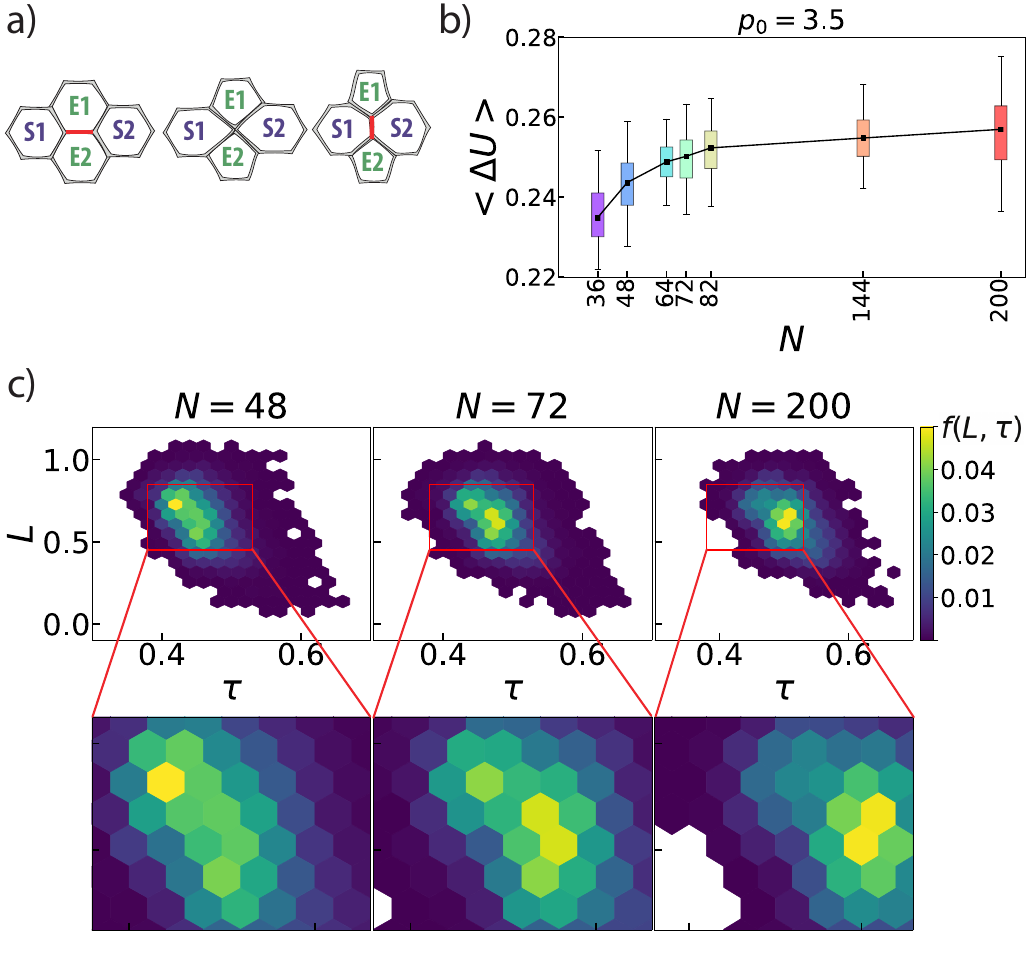}
\caption{a) Schematic of a T1 transition. Left: Cells E1 and E2 share initially an edge (orange line). Center: the edge is collapsed into a four-fold vertex. Right: A new edge is created between cells S1 and S2 (orange line). b) Whisker plots of $\Delta U$ versus $N$ at fixed $p_0=3.5$ and $k_A=1$. Boxes span from the first to the third quartile. Whiskers extend to the farthest data point within 1.5 times the interquartile range from the box. The black line indicates the median.
c) Distributions of edge lengths $L$ and tensions $\tau$ (top panels, magnification in bottom panels) at $N=48,72,200$, $p_0=3.5$, and $k_A=1$.}
\label{fig:3}
\end{figure}

\paragraph{\mdm{Curvature lowers the energy barrier for cellular rearrangements.}}
To gain mechanistic insights on the curvature-induced fluidization, we examine the mechanical response of the tissue to external perturbations. Under confluency, tissue fluidity is driven by cells' ability to rearrange via T1 transitions (Fig.\ref{fig:3}a)\cite{carvalho2018occluding,walck2014cell,irvine1994cell,tetley2019tissue,tada2012convergent,rauzi2020cell,spencer2017vertex}.
We hypothesize that tissue curvature impacts rigidity by affecting the energy $\Delta U$ required to execute a T1 transition, defined as the energy cost to collapse an edge into a fourfold vertex(SM\cite{suppl_mat}). We test this by studying how $\Delta U$ changes with curvature.

At fixed $p_0=3.5$ and $k_A=1$, $\Delta U$ decreases on average with increasing curvature (Fig.\ref{fig:3}b). This confirms that curvature favors cell intercalations energetically. To understand this behavior, we delve into the underlying mechanism of T1 transitions. During a T1, tissue energy is impacted by two effects(SM\cite{suppl_mat}): 1) the pure topological rearrangement of the connectivity network; and 2) the relaxation of the surrounding vertices to reach a new stable configuration.
We study these contributions separately. Interestingly, the relaxation and topological terms are positively correlated as $\Delta U_{\text{rel}} \sim \eta \Delta U_{\text{top}}$, with $\eta$ being independent of $N$(SM\cite{suppl_mat}).

\mdm{The functional dependence of $\Delta U_{\text{top}}$ can be predicted by developing a simple geometric ansatz. When an edge is collapsed, $\Delta U_{\text{top}}=\Delta U_{\text{per}} + \Delta U_{\text{area}}$ is defined only by the changes in area $\Delta U_{\text{area}}$ and perimeter $\Delta U_{\text{per}}$ of the four  cells involved in the T1. In a mean-field approximation, $\Delta U_{\text{per}}$ and $\Delta U_{\text{area}}$ can be derived analytically(SM\cite{suppl_mat}): 
\be
\Delta U_{\text{per}}=A L^2 + B L \tau_S + C L \tau; \quad \Delta U_{\text{area}}=k_A[\sin(2\pi/3)L^2]^2.
\label{delta_per}
\ee
$L$ and $\tau$ represent the initial length and tension of the collapsed edge. $\tau_S=p_{\text{S1}}+p_{\text{S2}}-2p_0$ expresses the deviation of the rescaled perimeters of cells $S_1$ and $S_2$ (Fig.\ref{fig:3}a) from $p_0$. $A$, $B$, and $C$ are numerical factors with values $A=1.085$, $B=1.29$, $C=-0.708$.
As such, $\Delta U_{\text{per}}$ depends only on the local properties of 1) the collapsed edge before the T1, and 2) the surrounding cells $S_1$ and $S_2$.
Similar to previous reports\cite{bi2014energy,bi2015density,sahu2020linear}, Eq.\ref{delta_per} includes a quadratic dependence of $\Delta U_{\text{per}}$ on $L$. Our model reveals also a linear dependence on the edge tension that is modulated by $L$. Hence, T1 transitions are energetically penalized on both longer and more tense edges.

By combining the relationship between $\Delta U_{\text{rel}}$ and $\Delta U_{\text{top}}$ with Eq.\ref{delta_per}, we recapitulate the behavior of $\Delta U$. $\Delta U$ shows no significant changes between edge configurations with same $(L,\tau,\tau_S)$ values but different $N$(SM\cite{suppl_mat}). At fixed $(\tau,\tau_S,p_0)$, $\Delta U$ collapses onto a single master curve that is independent of curvature and is well-described by our mean-field model(SM\cite{suppl_mat}). Such collapse occurs regardless of the $p_0$ value.
Our predictions hold also on flat surfaces(SM\cite{suppl_mat}), confirming the universality of Eq.\ref{delta_per}.

Our findings show that the T1's energetic cost depends solely on the local properties of the edge network. Surface curvature does not impact the T1's dependence on these properties. This suggests that the curvature-induced fluidization does not stem from local changes in the individual T1 mechanism, but rather from global changes in the edge network.}

\begin{figure}
\includegraphics[width=\linewidth]{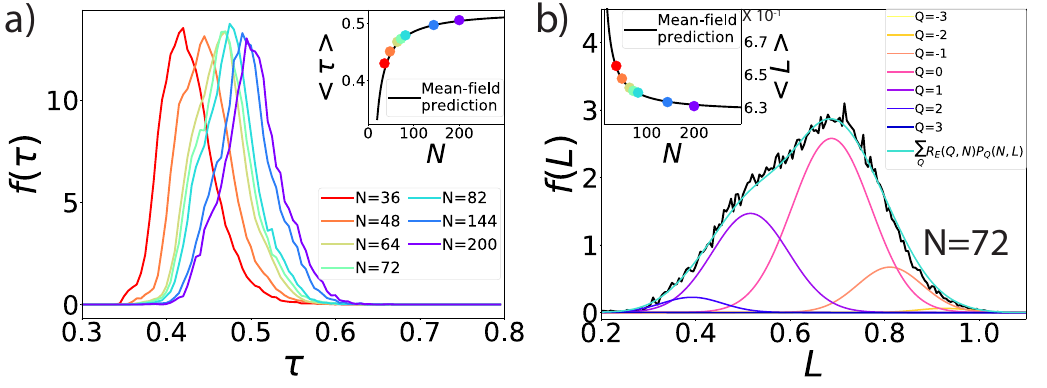}
\caption{a) Density distribution $f(\tau)$ of edge tensions for spheres with different $N$ at $p_0=3.5, k_A=1$. (Inset) Average edge tension $\langle \tau \rangle$ versus $N$. The solid line corresponds to Eq.\ref{tau-dep}, with $P_{\text{flat}}=3.76$ equal to the average perimeter at the largest $N=360$. b) Density distribution $f(L)$ of edge lengths at $N=72, p_0=3.5, k_A=1$ (black line). Colored solid lines correspond to Gaussian fits of the densities $f_Q(L)$. The turquoise line corresponds to the weighted sum of the Gaussian fits $\sum_Q{R_E(Q)f_Q(L)}$, where $R_E(Q)$ is the fraction of edges with given $Q$ over the total. (Inset) Average edge length $\langle L \rangle$ versus $N$ at $p_0=3.5$ and $k_A=1$. The solid line corresponds to Eq.\ref{L-dep}, with $P_{\text{flat}}=3.76$ equal to the average perimeter at the largest $N=360$.}
\label{fig:4}
\end{figure}

\paragraph{In the absence of cell motility, curvature of the jammed layer impacts the distributions of edge lengths and edge tensions.}To understand how changes in the cell-junctional network impact tissue rigidity, we first investigate the density distribution of edge length and tensions, $f(L,\tau)$, at $p_0=3.5, k_A=1$. Fig.\ref{fig:3}c confirms that curvature significantly alters $f(L,\tau)$. Consistent with our percolation analysis, the tension distribution $f(\tau)$ shifts towards larger $\tau$ values at higher $N$ (Fig.\ref{fig:4}a).
The distribution of edge lengths $f(L)$ exhibits a more intriguing behavior. On flat surfaces, $f(L)$ has been traditionally reported as a Gaussian\cite{bi2014energy}. On the spherical layer, however, $f(L)$ appears as the sum of multiple Gaussians $e^{(x-\mu)/\sigma}$ (Fig.\ref{fig:4}b). These Gaussians coincide with distinct contact topologies around the edge. Indeed, we find $(\mu,\sigma)$ to be parametric in the quantity $Q=6-Z_{\text{S1}}+6-Z_{\text{S2}}$, where $Z_{\text{S1}}$ and $Z_{\text{S2}}$ are the number of neighbors of cells S1 and S2\cite{bi2014energy} (Fig.\ref{fig:3}a). Edges are shorter at higher values of $Q$, corresponding to $\text{S1}, \text{S2}$ pairs with fewer neighbors. 
We observe similar behavior on flat surfaces(SM\cite{suppl_mat}), where a monotonic decrease of $\Delta U$ with $Q$ has been reported\cite{bi2014energy}. Our results offer mechanistic insights: shorter cell edges at larger $Q$ reduce the energetic cost of cellular rearrangements.

We analyze the edge length distribution $f_Q(L)$ at different $Q$ and radii of curvature. The average length $\mu_Q$ decreases with $N$ for each $Q$, while the ratio of edges with given $Q$ varies non-monotonically with $N$(SM\cite{suppl_mat}). This leads to an average increase of $L$ with increasing curvature (inset of Fig.\ref{fig:4}b).

The dependence of edge lengths and tensions on $N$ can be explained through geometric arguments. The perimeter $P$ of a spherical polygon is related to its equivalent $P_{\text{flat}}$ on the flat surface via the identity $P(N)=P_{\text{flat}}  \sqrt{1-1/N}$(SM\cite{suppl_mat} and\cite{hernandez2023finite}). By combining this scaling with the Euler's polyhedron formula, we can derive the average length $\langle L \rangle$ and tension $\langle \tau \rangle$:
\begin{align}
\langle L \rangle &\approx  \langle P\rangle/ \langle Z \rangle \approx \frac{\langle P_{\text{flat}} \rangle \sqrt{1-1/N}}{6-12/N};
\label{L-dep}\\
\langle \tau \rangle &\approx 2 \langle P \rangle- 2 P_0=\tau_{\text{flat}} + A_{\tau} \left(-1 + \sqrt{1-1/N} \right),
\label{tau-dep}
\end{align}
where $\tau_{\text{flat}}=2 \langle P_{\text{flat}} \rangle- 2 P_0$ and $A_{\tau}=2 \langle P_{\text{flat}} \rangle$. \mdm{$\langle \rangle$ denotes the average over all equilibrium configurations with given $(N,p_0,k_A)$ obtained via our energy minimization scheme.}
Both Eqs.\ref{L-dep}-\ref{tau-dep} recapitulate our simulations results (insets of Fig.\ref{fig:4}a-b and SM\cite{suppl_mat}). As such, curvature impacts the cell network's spatial (edge lengths) and mechanical (edge tensions) configuration in opposite ways. These effects, in turn, impact tissue rigidity in opposite ways: longer edges penalize cell rearrangements, while lower intercellular tensions favor cell intercalation. \mdm{Notably, Eq.\ref{tau-dep} provides information also on the density distribution $f(\tau_S)$. $f(\tau_S)$ exhibits the same trend as $f(\tau)$ when varying $N$(SM\cite{suppl_mat}). Since $\tau$ and $\tau_S$ are theoretically equivalent on average, Eq.\ref{tau-dep} correctly predicts the dependence of $\langle \tau_S \rangle$ on curvature(SM\cite{suppl_mat}).} 

Eq.\ref{L-dep} implies that the edge length distribution arises from the interplay of two constraints: the spherical constraint, which forces cells with the same area and higher curvature to have smaller perimeters, and the topological constraint, which forces the average number of cell neighbors to decrease with curvature. This \mdm{latter} effect dominates the behavior of $\langle L \rangle$, but is insufficient to explain alone its overall dependency on $N$. To test this, we compare our spherical VM with a spherical Voronoi tessellation, where varying $N$ changes only the number of topological defects \mdm{(defined as deviations from a regular hexagonal cellular network)} but not the surface curvature. Our simulations confirm that the decrease of $\langle L \rangle$ with $N$ due to the topological constraints of the Voronoi sphere does not recapitulate the behavior shown in Fig.\ref{fig:4}b(SM\cite{suppl_mat}).

The results presented so far have been derived at fixed area elasticity $k_A=1$. Given that the edge network at equilibrium depends on $k_A$ in principle, we ask whether different $k_A$ values affect the curvature-induced rigidity transition.
Based on our mean-field model, we expect $f(L)$ and $f(\tau)$, and hence $p_0^*$, to be independent on $k_A$(SM\cite{suppl_mat}). Analysis of the spherical VM at various $k_A$ confirms that, to the leading order, the impact of the area elasticity on the rigidity transition is negligible(SM\cite{suppl_mat}).

Cell packing also affects the edge network's configuration. Thus, we explore how the cell density $\rho$ impacts the rigidity of our spherical VM by relaxing the uniform density constraint $R=\sqrt{{N}/{4 \pi}}$. Analogous to the flat VM, increasing $\rho$ decreases the overall edge tension, shifting the onset of rigidity to lower $p_0$. For $\rho < 1$, the average edge tension collapses onto a single curve after rescaling $p_0$ by the average cell area. This suggests that the density-dependent changes in the cell tension network are trivially due to the corresponding changes in cell areas. Conversely, for $\rho > 1$, tissue rigidity is affected by additional non-trivial factors beyond the changes in cell area(SM\cite{suppl_mat}).

\paragraph{In the presence of cell motility, curvature promotes cellular migration.}
We further investigate how curvature affects tissue fluidity in the presence of cell motility. We consider the interplay between cell motility and substrate frictional drag\cite{tang2022collective}. In this description, the change rate of position $\vec{r}_i$ of the $i$-th vertex obeys the overdamped equation of motion $\Gamma d  \vec{r}_i/dt =\vec{F}_i^{\text{int}} + v_0 \hat{n}_i$, where $\Gamma$ is the frictional damping coefficient. $\vec{F}_i^{\text{int}}$ summarizes intercellular interactions mediated by cell-cell contacts and $v_0 \hat{n}_i$ represents cellular tractions applied on the substrate. Cell polarization vectors \mdm{$\hat{n}_i$} undergo rotational diffusion with rate $D_r$\cite{szabo2010collective,ladoux2016front,sknepnek2015active}(SM\cite{suppl_mat}).

\begin{figure}
\includegraphics[width=\linewidth]{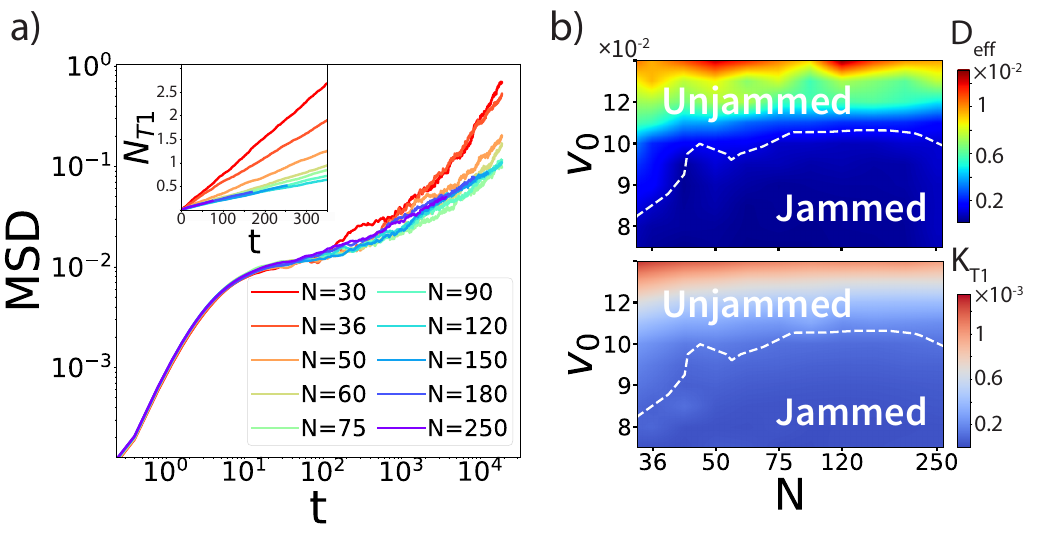}
\caption{a) Mean-square displacement on sphere varying $N$ at $p_0=3.65, k_A=1,v_0=0.09$. Inset: Cumulative count of T1 per junction vs. time. b) Heat maps of the diffusion coefficient $D_{\text{eff}}$ (upper panel) and T1 rate $k_{T1}$ (lower panel) as a function of $N$ and $v_0$ at fixed $p_0=3.65$ and $k_A=1$. The dashed line represents the motility-driven transition  $v_0^*(N)$, where $D_{\text{eff}}=0.001$\cite{bi2016motility,sahu2020small}.}
\label{fig:5}
\end{figure}

We first analyze the cell dynamics of the spherical layer at fixed $v_0=0.09$, $p_0=3.65$, and $k_A=1$. \mdm{This parameters' choice ensures that
the single-cell dynamics remains Brownian across all the investigated radii of curvature.} The average cell mean-square displacement $\langle \Delta r^2(t)\rangle$ displays three temporal regimes that are reminiscent of glassy dynamics (Fig.\ref{fig:5}a): an early-time ballistic regime, an intermediate switching regime that resembles a plateau, and a long-time diffusive regime.
For each $N$, we calculate the self-diffusion coefficient $D_{\text{eff}}$ in units of the free diffusion constant $D_0$ of an isolated cell. $D_{\text{eff}}$ is defined as $D_{\text{eff}}=D_s/D_0$, where $D_s=\lim_{t \to \infty} \langle \Delta r^2(t)\rangle/(4t)$ and $D_0=v_0^2/(2D_r)$.
$D_{\text{eff}}$ shows a strong dependence on curvature, exhibiting higher values at lower $N$(SM\cite{suppl_mat}). Similarly, the cumulative number $N_{T1}$ of T1 transitions per junction over time decreases with $N$ (Fig.\ref{fig:5}a inset).

We further measure $D_{\text{eff}}$ and the rate of T1 transitions $k_{T1}$ for several combinations of $v_0$ and $N$. At fixed $N$, there is a motility-driven solid-fluid transformation (Fig.\ref{fig:5}b). Notably, surface curvature modulates the cell motility threshold $v_0^*$ at which this transition occurs. 
$v_0^*$ increases non-monotonically with $N$, reaching an asymptotic plateau for $N>75$ where the effect of curvature becomes negligible (Fig.\ref{fig:5}b).
Hence, all other factors being equal, the epithelial layer is more fluid-like and unjammed when it resides upon more curved surfaces.

\paragraph{Discussion.} Our model indicates that increasing curvature induces fluidization and migration. This is not due to changes in the local mechanism of cell intercalation, which is independent of curvature, but rather to changes in the global structure of the cell-junction network, which becomes less \mdm{tensed} with increased curvature. Cell rearrangements become energetically advantageous at higher curvatures, resulting in cellular configurations that are more malleable and more migratory.

Part of our results is supported by recent works on the spherical VM\cite{sussman_2020,thomas2023shape,amiri2022random}. Different from these studies, however, our work is the first to 1) investigate the effect of curvature on both tissue's mechanical stability and migratory behavior and 2) identify its underlying mechanism. In doing so, we offer testable hypotheses about how tissue mechanics respond to substrate curvature. Recent experiments on lung alveolospheres showed that curvature enhances fluidity by modifying the cell packing distribution\cite{tang2022collective}. \mdm{Our study demonstrates that curvature affects the cell rearrangements rate by altering the cell-cell contact network}, providing a mechanistic framework for understanding these experimental results. Our results also suggest that measuring junctional tensions (e.g. via localised laser ablation\cite{fernandez2009myosin} or laser tweezers\cite{bambardekar2015direct}) can serve as a direct indicator of tissue rigidity on curved surfaces. This new predictive metric could be employed to distinguish between different mechanical phases in organoid models.

Our findings stem from the isotropic nature of the spherical VM and its constant cell density. This geometry is ideal for understanding the intrinsic effects of surface curvature and is relevant to various physiological and pathological contexts\cite{tang2022collective,trushko2020buckling,roshal2020crystal,brandstatter2023curvature}. Anisotropic curved geometries may show more complex behaviors, with non-trivial dependencies on area elasticity or asymmetries in cell tension. Our work lays the groundwork for future studies on these geometries, such as tubes, ellipsoids, and toroids. Further analysis is also necessary to elucidate the role of cell packing density on the rigidity of curved epithelia. 
Similarly, further research is needed to understand how our findings translate to a 3D model of curved epithelia\cite{rozman2020collective,PhysRevLett.130.108401}. For surface patches with varying principal curvatures, such as those observed in epithelial monolayers on wavy patterns\cite{luciano2021cell}, mechanical contributions from the basal and lateral surfaces create heterogeneous stress distributions. This can affect cell packing, motility, and unjamming transitions. Regions with higher curvature may fluidize more easily due to lower energy costs for cell rearrangements, while flatter regions may remain rigid. Future work should develop a 3D vertex model\cite{rozman2020collective,lou2022curvature} to provide insights into the full mechanics of curved epithelia, considering the apical, basal, and lateral surfaces.
\newline
\begin{acknowledgements}
The authors acknowledge the support of the Northeastern University Discovery Cluster. MDM acknowledges support from the National Heart Lung and Blood Institute (grant no. NHLBI K25HL168157). AD acknowledges support from the New Faculty SEED Grant from IIT Delhi, India and the Start-up Research Grant from SERB-DST, Govt of India (grant no. SRG/2023/000099). D.B. acknowledges support from the National Science Foundation (grant no. DMR-2046683 and PHY-2019745), the NIGMS of the National Institutes of Health (NIH) (grant no. R35GM15049), the Alfred P. Sloan Foundation, and the Human Frontier Science Program (Ref.-No.:  RGP0007/2022). JJF acknowledges support from the National Heart Lung and Blood Institute (grant no. NHLBI 1R01HL148152 and T32HL007118).
\end{acknowledgements}

\bibliography{bib_UJT_sphere}
\clearpage

\newpage
\beginsupplement

\begin{center}
\textbf{\large Supplemental Materials for Epithelial layer fluidization by curvature-induced unjamming}
\end{center}

\section{Simulation details}
To initialize the simulation, the vertices positions are generated via a Voronoi tesselation of a random set of points uniformly distributed on the sphere. A two-step minimization procedure is implemented to derive tissue equilibrium configurations. Using the open-source software Surface Evolver\cite{brakke1992surface}, we first minimize Eq.\ref{energy} at $p_0^{\text{in}}=3.695$ through a combination of the first-order conjugate-gradient and second-order Newton's method\cite{kelley1999iterative}. In this step, the contact topology is updated via T1s transitions\cite{bi2014energy,das2021controlled,bi2015density,spencer2017vertex,weaire1999physics}. T1 transitions are performed in sequence on edges with lengths $L<L_{\text{min}}=\{0.01,0.02,0.05,0.1,0.2,0.3\}$ and this process is iterated four times. At each T1 transition, cells swap neighbors and the global edge network is relaxed via energy minimization. After this step, we increase the target shape index to $p_0=4$ and progressively decrease it by intervals of $\Delta p_0=0.005$. At each intermediate $p_0$, the system's energy is minimized by allowing only vertices rearrangements. In this step, the contact topology remains fixed.

\mdm{While the conclusions of the paper are not dependent on the specific minimization procedure used, the sequence of T1 thresholds is chosen to ensure meaningful comparisons across different model parameters and spherical curvatures. Performing a sequence of T1 transitions at progressively larger edge-length thresholds provides a reliable means of gradually “annealing” the cell-contact topology to a stable distribution, independent of surface curvature. This yields a more stable contact network topology and avoids trapping the system in metastable states, which can occur when all T1 transitions are performed below a single, large threshold at once. Hence, our approach allows isolating the effect of curvature and cell contact network topology.}

Overall, we generate a set of equilibrium configurations with the same contact topology and model parameters in the range of $p_0=3.5-4$ and $N=36-360$. We analyze these ground states for different values of the area elasticity $k_A=\{0.05,0.1,0.5,1\}$. For each choice of the parameters $(N,p_0,k_A)$, $200$ initial states are randomly generated and equilibrated. The code used to simulate our spherical VM, along with the relevant input files, is available on GitHub at \url{https://github.com/marghe-88/spherical_VM.git}.

\section{Analysis of percolation of the tension network}
For each of the $200$ equilibrium configurations with given $(N,p_0,k_A)$, we extract the subset of edges with \mdm{$\tau>10^{-20}$}. These edges define the tissue's tension network. In the main text, snapshots of the tension network show that the network changes its size and connectivity by varying $p_0$ and $N$. To investigate the tension percolation state at different $(p_0,N,k_A)$, we compute the largest connected component (LCC) of each network and check if the LCC has percolated. We define a percolating LCC as a cluster in which the maximum Cartesian distance between each vertex pair is larger than $2R-\sqrt{\bar{A}}$, with $R$ the sphere radius and $\bar{A}$ the average cell area. In this way, a percolating LCC spans at least half of the sphere. For a given choice of $(N,p_0,k_A)$, the tension percolation probability $F_{\text{perc}}(N,p_0,k_A)$ is defined as the fraction of tension networks that has percolated. We repeat this process for multiple $p_0$, $N$, and $k_A$.

We further compute the percolation threshold $p_0^*(N,k_A)$. $p_0^*(N,k_A)$ is defined such that $F_{\text{perc}}(N,p_0^*,k_A)=0.5$, i.e. half of the $200$ random seeds have percolated. After shifting $F_{\text{perc}}$ by its relative $p_0^*$, we calculate the scaling exponent $\nu$ as the value at which $F_{\text{perc}}(N,p_0-p_0^*(N),k_A)$ collapses onto a single curve.

We have shown in the main text that $p_0^*$, and hence the onset of rigidity, increases with $N$. This dependence can be understood within the framework of geometric compatibility. Previous studies on under-constrained spring networks have shown that the relationship between isotropic strain and spring lengths can be expressed as a minimal-length function\cite{Lee_Merkel_sm,merkel2019minimal}. Based on this function, a non-dimensional control parameter $l_0$ exists that marks the rigidity transition and depends solely on the geometry of the network. Applying isotropic strain $\epsilon$ rescales $l_0$, leading to a rescaling of the dimensional rest spring length $L_0$:
\be
L_0 \propto l_0 \left(1+\epsilon\right).
\label{strain}
\ee
In our spherical VM, the effect of curvature is analogous to the one of isotropic strain: edge perimeters are uniformly stretched with increasing $N$. By replacing strain with curvature and the rest spring length with the target shape index in Eq.\ref{strain}, \mdm{we can use the theoretical predictions of Eq.\ref{spher_cap}} and derive the dependence of $p_0^*$ with $N$:
\be
p_0^*(N)=p^*_{0_{\text{flat}}} \sqrt{1-1/N},
\label{strain_p0}
\ee
where $p^*_{0_{\text{flat}}}$ represents the onset of rigidity in the flat VM. Notably, Eq.\ref{strain_p0} recapitulates the shift of the tension percolation threshold $p_0^*(N)$, as shown in the inset of Fig.\ref{fig:2}c.

\section{Calculation of energy barrier associated with a T1 transition}
\begin{figure*}
\includegraphics[width=\linewidth]{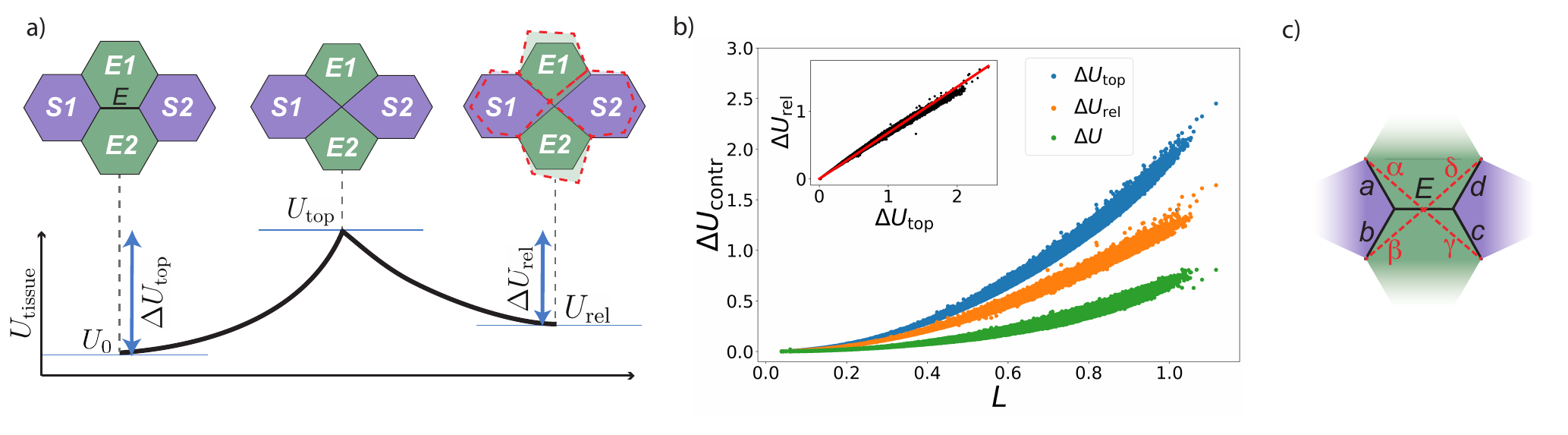}
\caption{a) Schematic of the tissue's energy profile associated with the topological and relaxation contributions. In the initial state, two adjacent cells (colored in green) share an edge $E$ (left panel). When $E$ is collapsed into a fourfold vertex while keeping the other vertices fixed, the tissue's energy increases to $U_{\text{top}}$ (central panel). After the topological collapse, the layer rearranges the surrounding vertices and relaxes into a new state with lower energy $U_{\text{rel}}$ (right panel). This rearrangement involves vertices belonging to other cells beyond the four cells involved in the T1 (not shown in the figure). b) T1's energy contributions versus $L$. At each T1, $\Delta U_{\text{top}}$ is always larger than $\Delta U_{\text{rel}}$. $\Delta U_{\text{rel}}$ is well fitted by the linear function $\Delta U_{\text{rel}} = \eta \Delta U_{\text{top}}$ (red line in the inset). c) \mdm{Schematic of the T1 topological collapse in our mean-field model. The geometric collapse of the edge $E$ into a fourfold vertex affects only the four edges $(a,b,c,d)$ sharing a vertex with $E$. We assume these edges to have equal initial lengths $L(a)=L(b)=L(c)=L(d)=L(E)=L$ and form regular hexagon angles with $E$, while the remaining cell edges can vary in number and length.}}
\label{supp_fig:1}
\end{figure*}

Under confluent conditions, cells swap neighbors by reorienting their local connectivity network. This intercalation process is referred to as a T1 transition. During a T1 transition, an existing edge is collapsed into a fourfold vertex to create a new junction in the perpendicular orientation (Fig.\ref{fig:3}a). For each of the $200$ states with given $(N,p_0)$ and $k_A=1$, we perform a T1 transition on each edge of the spherical layer. \mdm{We implement the T1 transition as a two-step process. In the first step, the selected edge is geometrically collapsed to a very short length ($10^{-6}$). We define $\Delta U_{\text{top}}$ as the increase in the tissue’s total mechanical energy, as defined in Eq.\ref{energy}, resulting from this topological collapse (central panel of Supp. Fig.\ref{supp_fig:1}a). In the second step, we equilibrate the system by keeping the fourfold vertex fixed and allowing the other vertices to rearrange. We define $\Delta U_{\text{rel}}$ as the decrease in the system's mechanical energy resulting from this relaxation (right panel of Supp. Fig.\ref{supp_fig:1}a). $\Delta U_{\text{rel}}$ accounts for both local effects from the vertices motion of the four cells surrounding the collapsed edge and non-local effects from those farther away. The total energy barrier $\Delta U$ associated with the T1 transition is calculated as $\Delta U=\Delta U_{\text{top}}-\Delta U_{\text{rel}}$. Similar to previous studies\cite{bi2014energy}, $\Delta U$ corresponds to the global energetic response of the tissue to the edge collapse.}
We analyze the statistics of $\Delta U_{\text{top}}$ and $\Delta U_{\text{rel}}$ separately. We find that $\Delta U_{\text{top}} > \Delta U_{\text{rel}}$ in every T1 transition (Supp. Fig.\ref{supp_fig:1}b) and the two contributions are linearly correlated $\Delta U_{\text{rel}} = \eta \Delta U_{\text{top}}$, where $\eta=0.68$ (inset of Supp. Fig.\ref{supp_fig:1}b).

\section{Mean-field model of T1 transitions in the spherical VM}
To understand the functional dependence of $\Delta U$, we develop a mean-field model (Supp. Fig.\ref{supp_fig:1}c). The pure topological collapse affects only the collapsed edge $E$ and the four edges sharing vertices with $E$ (edges $a,b,c,d$ in Supp. Fig.\ref{supp_fig:1}c). \mdm{We consider a mean-field approximation wherein these four edges 1) have initial lengths $L(a)=L(b)=L(c)=L(d)=L(E)=L$, and 2) form regular hexagon angles with $E$. The cell edges that are not involved in the topological collapse can vary in number and length (Supp. Fig.\ref{supp_fig:1}c). When $E$ is collapsed into a fourfold vertex, the new length of the four edges becomes:
\be
\label{L_final}
L(\alpha)=\sqrt{L(a)^2+\frac{L^2}{4}-L(a)L\cos(k)},
\ee
where $k$ is the angle between $a$ and $E$. Since $k=\frac{4}{6}\pi$ in regular hexagons and $L(a)=L$, Eq.\ref{L_final} reduces to:
\be
L(\alpha)=L(\beta)=L(\gamma)=L(\delta)=L\sqrt{\frac{5}{4}-\cos\left(\frac{4}{6}\pi\right)}.
\ee

By definition, $\Delta U_{\text{top}}=\Delta U_{\text{per}}+\Delta U_{\text{area}}$, where:
\be \begin{split}
\Delta U_{\text{per}} &= \sum_{i\in\{\text{S1},\text{S2},\text{E1},\text{E2}\}}{(p_i^{\text{fin}}-p_0)^2 - (p_i^{\text{in}}-p_0)^2 } \\
&=\sum_{i\in\{\text{S1},\text{S2},\text{E1},\text{E2}\}}{\Delta p_i^2+2p_i^{\text{in}}\Delta p_i-2\Delta p_i p_0}.
\label{delta_per_eq}
\end{split} \ee
$\Delta p_i=p_i^{\text{fin}}-p_i^{\text{in}}$, with $p_i^{\text{in}}$ and $p_i^{\text{fin}}$ the perimeter of the i-th cell before and after the topological collapse, respectively. $\Delta p_i$ depends geometrically on $L$ as follows:
\begin{align}
\label{delta_per_S}
&\Delta p_{\{\text{S1},\text{S2}\}} = -2L+2L\sqrt{\frac{5}{4}-\cos\left(\frac{4}{6}\pi\right)}\\
\label{delta_per_E}
&\Delta p_{\{\text{E1},\text{E2}\}} = -3L+2L\sqrt{\frac{5}{4}-\cos\left(\frac{4}{6}\pi\right)}.
\end{align}
Combining Eq.\ref{delta_per_eq} with Eqs.~\ref{delta_per_S} and \ref{delta_per_E}, we arrive to the final equation:
\be
\label{final_eq_functional}
\Delta U_{\text{per}}=A L^2 + B L \tau_S + C L \tau,
\ee
where $\tau=p_{\text{E1}}^{\text{in}}+p_{\text{E2}}^{\text{in}}-2p_0$ and $\tau_S=p_{\text{S1}}^{\text{in}}+p_{\text{S2}}^{\text{in}}-2p_0$. $\tau$ and $\tau_S$ quantify how much the $(E_1,E_2)$ and $(S_1,S_2)$ cell perimeters (Supp. Fig.\ref{supp_fig:1}a) differ from $p_0$, respectively. $\tau$ coincides with the initial tension of the collapsed edge $E$. The coefficients $(A,B,C)$ are uniquely determined by the angle $k$:
\be
\begin{split}
A &= 8(G-1)^2 + 2(2G-3)^2 = 1.085\\
B &= 4(G-1) = 1.29\\
C &= 2(2G-3) = -0.708,
\end{split}
\ee
with $G=\sqrt{\frac{5}{4}-\cos(k)}$. Similarly, $\Delta U_{\text{area}}$ can be written as:
\begin{align}
\Delta U_{\text{area}} &= k_A \sum_{i\in\{\text{S1},\text{S2},\text{E1},\text{E2}\}}{(a_i^{\text{fin}}-1)^2 - (a_i^{\text{in}}-1)^2}\\
\label{delta_ar_eq}
&=k_A \sum_{i\in\{\text{S1},\text{S2},\text{E1},\text{E2}\}} \Delta a_i^2+2a_i^{\text{in}}\Delta a_i-2\Delta a_i,
\end{align}
where $\Delta a_i = a_i^{\text{fin}} - a_i^{\text{in}}$. The terms $a_i^{\text{in}}$ and $a_i^{\text{fin}}$ represent the area of the i-th cell before and after the topological collapse, respectively. For each of the four cells involved in the T1, the area difference $\Delta a_i$ depends solely on the area of the triangle $T_a$ formed by the three segments $\{a,\alpha, E/2\}$. Specifically:
\begin{align}
\label{delta_ar_S}
&\Delta a_{\text{S1},\text{S2}} = 2T_a\\
\label{delta_ar_E}
&\Delta a_{\text{E1},\text{E2}} = - 2T_a.
\end{align}
Combining Eq.\ref{delta_ar_eq} with Eqs.~\ref{delta_ar_S} and \ref{delta_ar_E}, $\Delta U_{\text{area}}$ reduces to:
\be
\Delta U_{\text{area}} = k_A 16 T_a^2=k_A \sin^2\left(\frac{4}{6}\pi\right)L^4.
\ee
Given that $\Delta U_{\text{rel}} = \eta \Delta U_{\text{top}}$ and $\Delta U = \Delta U_{\text{top}} - \Delta U_{\text{rel}}$, the final functional dependence of $\Delta U$ can be written as:
\be
\label{final_master_eq}
\Delta U(L,\tau,\tau_S) = \left( 1-\eta \right) \left( A L^2 + B L \tau_S + C L \tau + k_A \sin^2 \left( \frac{4}{6} \pi \right) L^4 \right)
\ee
It follows that the T1's energy barrier depends solely on the local properties $(L,\tau,\tau_S)$ of the edge network and is independent of the surface curvature.}

\begin{figure*}
\includegraphics[width=\linewidth]{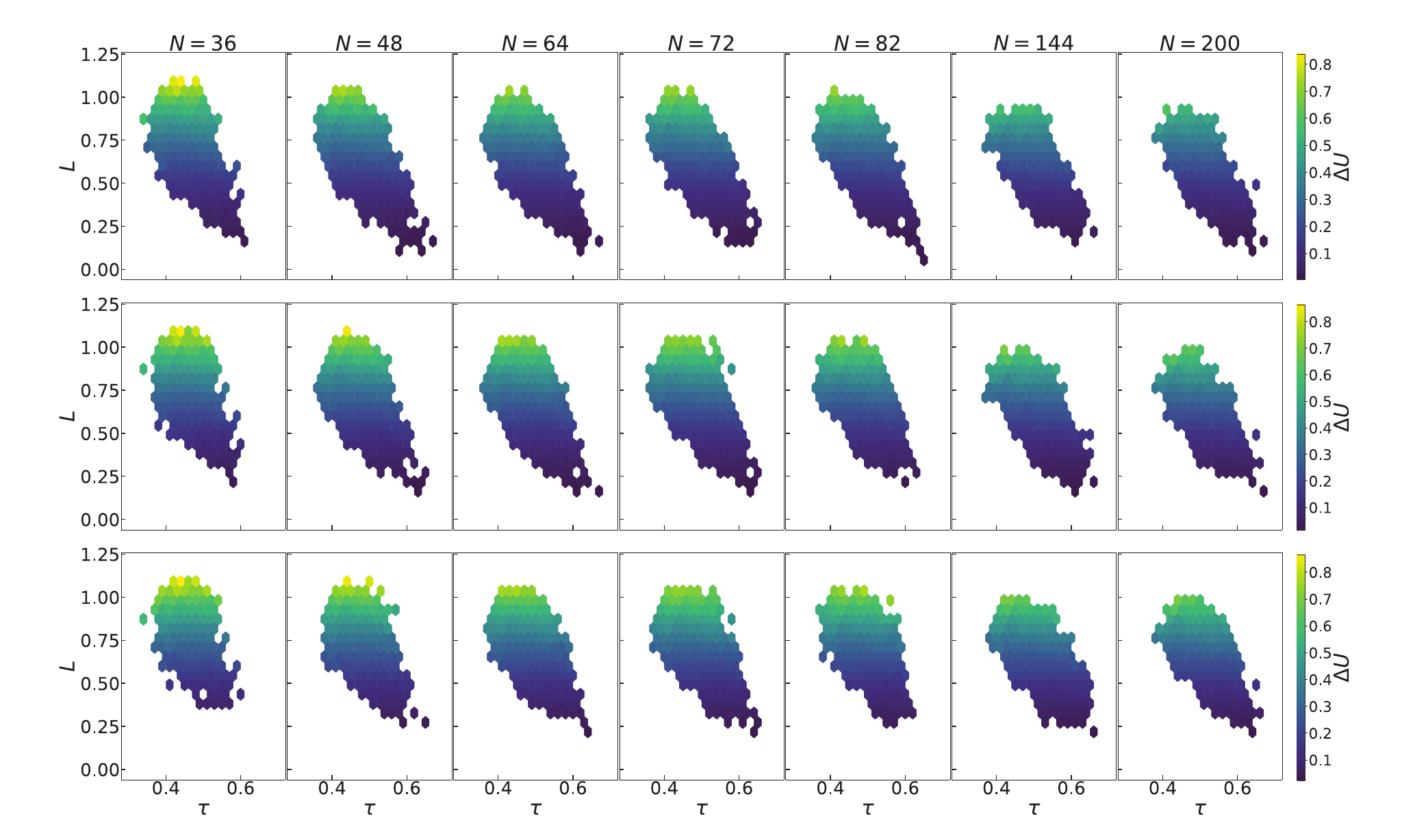}
\caption{\mdm{Distributions of energy barriers $\Delta U(L,\tau,\tau_S)$ in spherical VMs with different $N$ for three combinations of $(\tau_S,p_0,k_A)$: $\{0.48,3.5,1\}$ (top panels), $\{0.5,3.5,1\}$ (central panels), and $\{0.52,3.5,1\}$ (bottom panels). For each $(\tau_S,p_0,k_A)$ combination, the $\Delta U$ associated with edges with same initial length and tension show no significant changes with curvature. The specific $\tau_S$ values in the figure were selected to maximize the edge count for visualization purposes.}}
\label{supp_fig_Round2:2}
\end{figure*}

\begin{figure*}
\includegraphics[width=\linewidth]{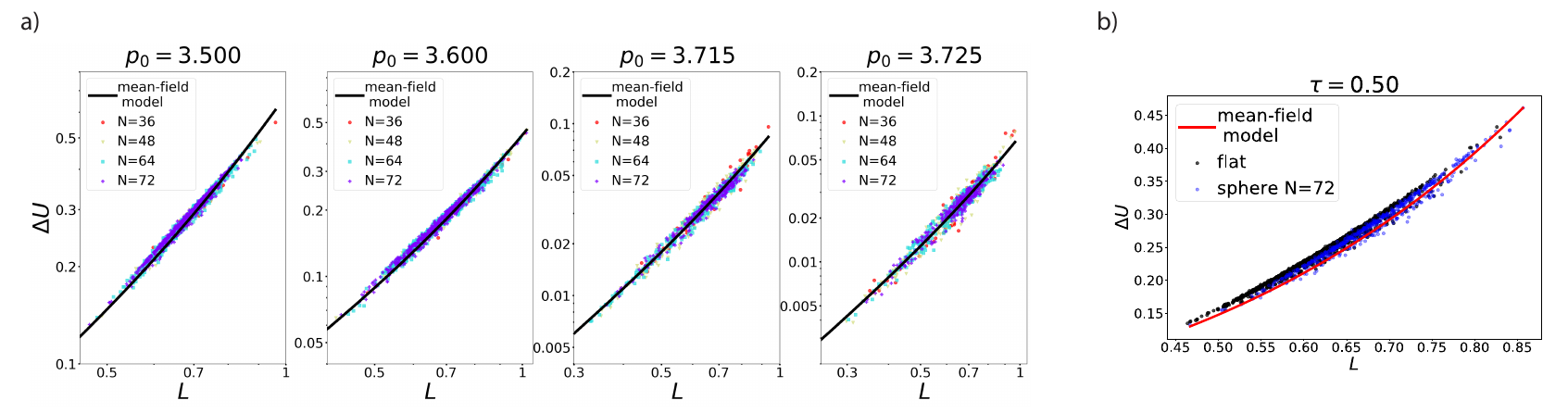}
\caption{\mdm{a) T1's energy barriers $\Delta U$ versus $L$ for four different combinations of $(p_0,\tau,\tau_S)$: $p_0=\{3.5,3.6,3.715,3.725\}$ and corresponding $\tau=\tau_S=\{0.5, 0.30, 0.04, 0.025\}$. $k_A$ is fixed at $k_A=1$. Different colors correspond to spheres with different number of cells $N=\{36,48,64,72\}$. The black solid line corresponds to Eq.\ref{final_master_eq}. While the spherical VM exhibits a range of $(L,\tau,\tau_S)$ distributions for each $p_0$, the specific $(\tau,\tau_S)$ values in the figure were selected to maximize the edge count for visualization purposes. b) T1's energy barriers $\Delta U$ versus $L$ at fixed $\tau=\tau_S=0.5$, $p_0=3.5$, and $k_A=1$ in the flat VM ($N=300$, black dots) and spherical VM ($N=72$, blue dots). The red solid line corresponds to Eq.\ref{final_master_eq}.}
}
\label{new_supp_fig_Round2:3}
\end{figure*}

\mdm{
To validate our predictions, we analyze the distribution of $\Delta U$ across various curvatures. Supp. Fig.\ref{supp_fig_Round2:2} shows $\Delta U$ as a function of $(L,\tau)$ for T1 transitions with fixed $\tau_S=\{0.48,0.5,0.52\}$ ($p_0=3.5$ and $k_A=1$). Similar to flat surfaces\cite{bi2014energy,bi2015density,das2021controlled,sahu2020linear}, $\Delta U$ increases with $L$. In agreement with our mean-field model, for each $\tau_S$, $\Delta U(L,\tau)$ shows no significant changes between edges with same $(L,\tau)$ values but different $N$.}
\mdm{Supp. Fig.\ref{new_supp_fig_Round2:3}a further confirms the functional dependence predicted in Eq.\ref{final_master_eq}. It displays the log-log distribution of $\Delta U$ at various $N$ as a function of $L$ for four combinations of $(p_0,\tau,\tau_S)$: $p_0=\{3.5,3.6,3.715,3.725\}$ and corresponding $\tau=\tau_S=\{0.5, 0.30, 0.04, 0.025\}$ ($k_A=1$). The $(p_0,\tau,\tau_S)$ values were intentionally chosen to ensure the system’s rigidity and maximize the edges count across all analyzed $N$ values. For each $(p_0,\tau,\tau_S)$, $\Delta U(L)$ collapses onto a single master curve independently of the radius of curvature. Notably, this curve is well-described by the mean-field model's predictions in Eq.\ref{final_master_eq}.}

\mdm{Since Eq.\ref{final_master_eq} is independent of the surface topology, we demonstrate its validity also on flat surfaces. We simulate a flat VM with $N=300$ cells at $p_0=3.5$ and $k_A=1$. We compute the T1's energy barriers $\Delta U$ using the same energy minimization scheme of the spherical VM. Supp. Fig.\ref{new_supp_fig_Round2:3}b shows the distribution of $\Delta U(L)$ at fixed $\tau=\tau_S=0.5$ for the flat VM and spherical VM ($N=72$, $p_0=3.5$, and $k_A=1$). Both distributions collapse onto the same master curve, which follows the behavior of Eq.\ref{final_master_eq}. As such, the functional dependence of $\Delta U(L,\tau,\tau_S)$ predicted by our mean-field model represents a universal law that is independent of the surface topology.}

\section{Distribution of edge lengths in the spherical and flat VM}
\begin{figure}
\includegraphics[width=\linewidth]{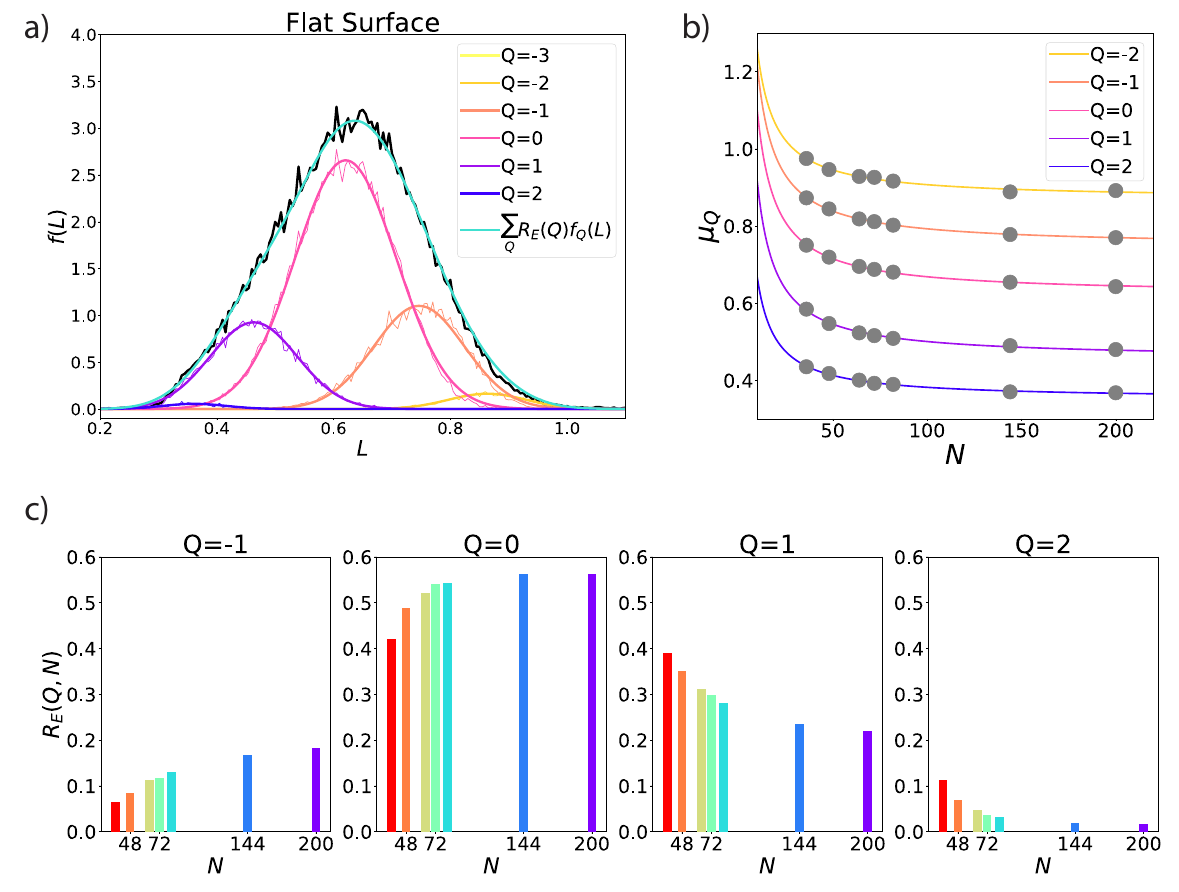}
\caption{a) Density distribution $f(L)$ of edge lengths in a flat VM with $N=300$, $p_0=3.5$, and $k_A=1$ (black line). Each colored line represents the density distribution $f_Q(L)$ of edge lengths with the same $Q$. For each $f_Q(L)$, the smooth line corresponds to its Gaussian fit $f_Q(L)=e^{(L-\mu_Q)/\sigma_Q}$. The turquoise line corresponds to the weighted sum of the Gaussian fits $\sum_Q{R_E(Q)f_Q(L)}$. $R_E(Q)$ is the fraction of edges with given $Q$ over the total. b) Behavior of the average length $\mu_Q$ versus $N$ in the spherical VM at $p_0=3.5$ and $k_A=1$. c) Dependence on $N$ of the fraction of edges $R_E(Q,N)$ with given $Q$ in the spherical VM ($p_0=3.5$, $k_A=1$).}
\label{supp_fig:3}
\end{figure}

Fig.\ref{fig:4}b shows that edge lengths are stratified by a topological parameter $Q$ that expresses the local connectivity around the edge.
We compare $f(L)$ in the flat VM with $N=300$ and spherical VMs with different $N$ at fixed $p_0=3.5$ and $k_A=1$. \mdm{In both the flat and spherical VM, 1) the edge length distribution $f(L)$ appears as the sum of multiple Gaussian parametric in $Q$, and 2) the mean $\mu_Q$ of each Gaussian decreases with increasing $Q$ (Fig.\ref{fig:4}b and Supp. Fig.\ref{supp_fig:3}a).  Hence, these two aspects of $f(L)$ are independent of the surface topology. When comparing $\mu_Q$ for a fixed $Q$ across different $N$ in the spherical VM, the average length $\mu_Q$ increases with increasing curvature (Supp. Fig.\ref{supp_fig:3}b). Curvature also alters the distribution of $Q$. By increasing $N$, the proportion of edges with $Q>0$ decreases while the one with $Q\le0$ increases (Supp. Fig.\ref{supp_fig:3}c). This hints at a key role of the hexagonal configuration (where $Z_{\text{S1}}=Z_{\text{S2}}=6$ and $Q=0$) not only in the layer's energetic stability\cite{staple2010mechanics,bi2015density} but also in the contact topology of curved surfaces.}

\section{Dependence of the perimeter of a spherical polygon on curvature}
To understand the dependence of the edge lengths and tensions on the radius of curvature $R$ (or equivalently on the number of cells $N$), we propose a simple mean-field model. By construction, the average solid angle $\Omega$ subtended by an individual cell depends on $N$ as: 
\be
\Omega = 4 \pi /N = R^{-2}
\ee
We first consider the simple case where the angle is subtended by a spherical cap of polar angle $\phi$. Although cells have polygonal boundaries rather than circular, we will show that this difference enters only as a different factor in our final predictions. The relationship between $\Omega$ and $\phi$ is given by:
\be
\Omega = 4\pi \sin^2\left(\frac{\phi}{2}\right)
\label{solid_az_angle}          
\ee
Additionally, the rim of the spherical cap has a perimeter given by: 
\be
P_{\text{circle}} = 2 \pi R \sin \phi = 2 \sqrt{\pi} \sqrt{1-\frac{1}{N}}
\label{p_cap}
\ee
In the limit of large $N$, $P_{\text{circle}}$ approaches the flat value of $\sqrt{4 \pi} \approx 3.544$. 

Eq.\ref{p_cap} can be generalized to an arbitrary polygon. By substituting the factor $\sqrt{4 \pi}$ with the term $P_{\text{flat}}$ representing the polygon's perimeter on flat surfaces, we can extrapolate the curvature dependence of the perimeter $P(N)$ of a polygon on the sphere:  
\be
P(N) = P_{\text{flat}}  \sqrt{1-\frac{1}{N}} =  P_{\text{flat}}  \sqrt{1-\frac{1}{4 \pi R^2}}.
\label{spher_cap}
\ee
$P_N$ decreases with decreasing $N$, and hence increasing curvature. 

\begin{figure}
\includegraphics[width=\linewidth]{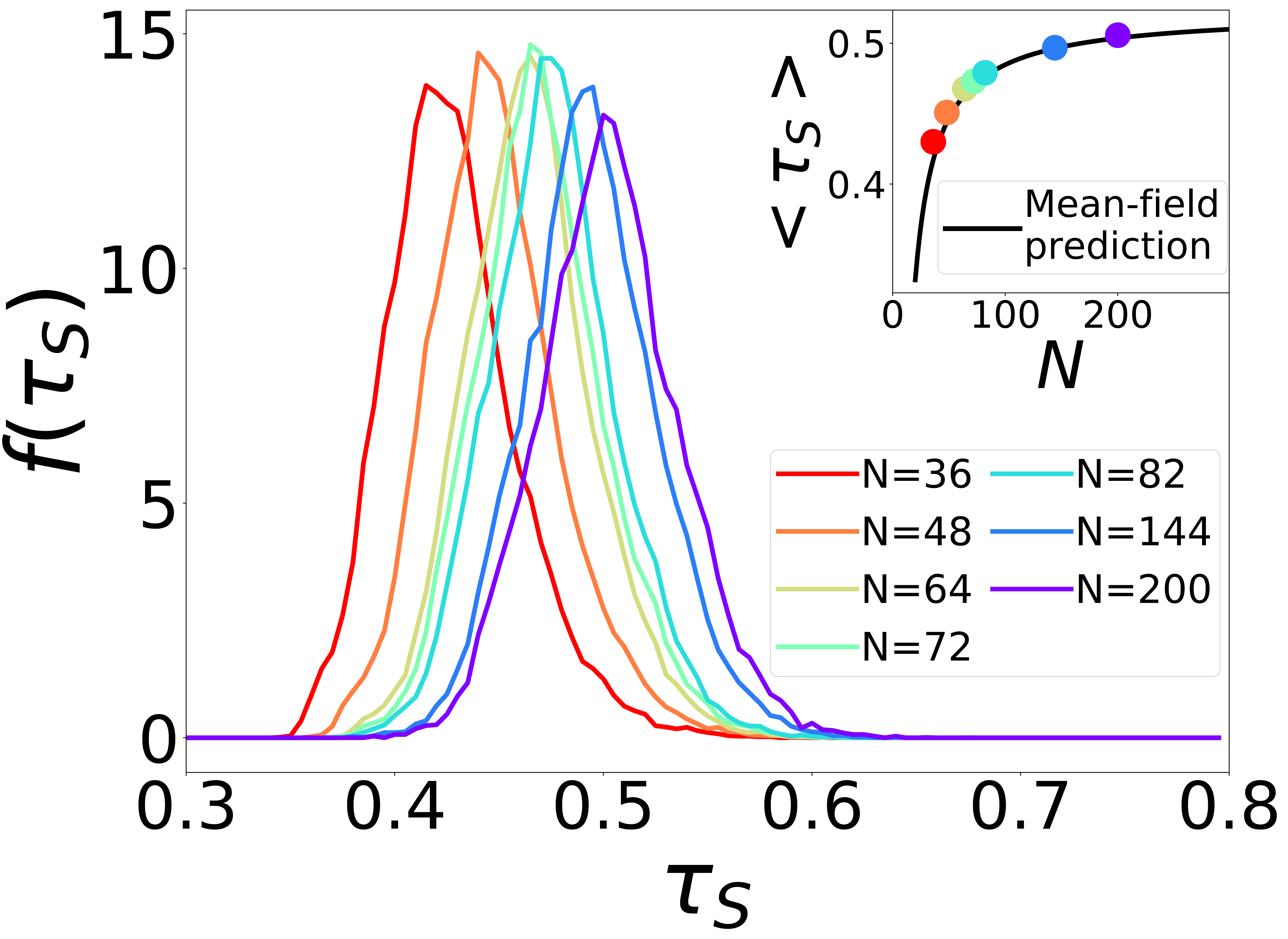}
\caption{\mdm{Density distribution $f(\tau_S)$ for spheres with different $N$ at $p_0=3.5, k_A=1$. (Inset) Average $\langle \tau_S \rangle$ versus $N$. The solid line corresponds to Eq.\ref{supp:tau_average}, with $P_{\text{flat}}=3.76$ equal to the average perimeter at the largest $N=360$.}
}
\label{new_supp_fig:5_Round2}
\end{figure}

Based on Euler's polyhedron formula, the number of vertices $V$, edges $H$, and cells $N$ tiling a spherical surface obey the relationship:
\be
N+V-H=2.
\label{Euler}
\ee
\mdm{Since the Voronoi tessellation imposes that each vertex connects with three edges}, $V=2H/3$. The number of edges can be written as $H=N \langle Z \rangle /2$, with $\langle Z \rangle$ the average number of cell neighbors. If we substitute these two relationships into Eq.\ref{Euler}, we obtain the dependence of $\langle Z \rangle$ from the number of cells $N$:
\be
H = 3 (N-2) = \frac{1}{2} N \langle Z \rangle \Rightarrow \langle Z \rangle = 6 - \frac{12}{N}
\label{Z_average}
\ee
By combining Eq.\ref{spher_cap} with Eq.\ref{Z_average}, we can derive the overall dependence of the average edge length and tension on $N$:
\be
\langle L \rangle  \approx  \frac{\langle P \rangle} {\langle Z \rangle} \approx \frac{\langle P_{\text{flat}} \rangle \sqrt{1-1/N}}{6-12/N}
\label{supp:L_average}
\ee
\mdm{
\be
\begin{split}
\langle \tau \rangle \approx 2 \langle P \rangle- 2 P_0 & = 2\langle P_{\text{flat}} \rangle\sqrt{1-\frac{1}{N}} - 2P_0\\
&=2\langle P_{\text{flat}} \rangle\sqrt{1-\frac{1}{N}} - 2P_0 + 2\langle P_{\text{flat}} \rangle - 2\langle P_{\text{flat}} \rangle\\
&=A_{\tau} \left(-1 + \sqrt{1-1/N} \right) + \tau_{\text{flat}}
\end{split}
\label{supp:tau_average}
\ee}
where $\tau_{\text{flat}}=2 \langle P_{\text{flat}} \rangle- 2 P_0$ and $A_{\tau}=2 \langle P_{\text{flat}} \rangle$.
\mdm{Both Eqs.\ref{supp:L_average}-\ref{supp:tau_average} accurately predict the average dependence of edge tensions $\tau$ and lengths $L$ on $N$. Notably, the distribution $f(\tau_S)$ exhibits a behavior similar to $f(\tau)$, shifting towards lower $\tau_S$ values when increasing curvature (Supp. Fig.\ref{new_supp_fig:5_Round2}). Since $\tau$ and $\tau_S$ are equivalent on average, $\langle \tau_S \rangle$ is also well-described by Eq.\ref{supp:tau_average} (inset of Supp. Fig.\ref{new_supp_fig:5_Round2}).}

\begin{figure}
\includegraphics[width=\linewidth]{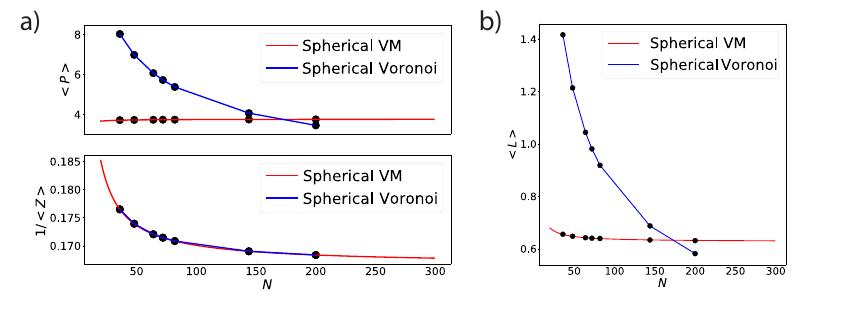}
\caption{
a) Average cell perimeter $\langle P \rangle$ (top) and number of cell neighbors $\langle Z \rangle$ (bottom) as a function of $N$ in the spherical Voronoi tessellation with fixed radius $R=3.45$ (blue lines) and spherical VM (red lines, $p_0=3.5$ and $k_A=1$). b) Average edge length $\langle L \rangle$ as a function of $N$ in the spherical Voronoi tessellation (blue line) and spherical VM (red line, $p_0=3.5$ and $k_A=1$). Solid lines serve as guides to the eye.
}
\label{new_supp_fig:4}
\end{figure}

Eq.\ref{supp:L_average} suggests that the dependence of the average edge length on $N$ results from the interplay of two factors: the spherical constraint, which forces cells with different surface curvature to scale their perimeter as $P_{\text{flat}}\sqrt{1-1/N}$, and the topological constraint, which forces the average number of cell neighbors to scale as $6-12/N$. Since $\langle L\rangle$ decreases with $N$, the topological constraint dominates this behavior. However, considering solely topological defects is insufficient for explaining this decrease. To illustrate this, we compare the edge length distribution of our spherical VM ($p_0=3.5$, $k_A=1$) with that of a spherical Voronoi tessellation with same $N$. \mdm{Unlike the spherical VM, in the spherical Voronoi tessellation: 1) the radius of curvature is fixed and does not vary with the number of cells and 2) cell network configurations are obtained through the Voronoi tessellation and no energy minimization procedure is performed.} This system represents a reference control where varying $N$ only varies the number of topological defects without altering the surface curvature. Supp. Fig.\ref{new_supp_fig:4}a shows that the spherical VM and Voronoi tessellation exhibit the same scaling of $\langle Z\rangle$.
Yet, the Voronoi tessellation displays a notably steeper decrease in $\langle L\rangle$ compared to the VM (Supp. Fig.\ref{new_supp_fig:4}b). In the Voronoi tessellation, cell perimeter is not bound by the spherical constraint, thus amplifying the impact of topological defects on $\langle L\rangle$  with respect to the VM (Supp. Fig.\ref{new_supp_fig:4}a). This confirms that varying $N$ in the spherical VM affects edge lengths in a complex manner, resulting from the combination of intrinsic curvature and topological effects.

\section{Influence of area elasticity and cellular density on the curvature-induced rigidity transition}
Since the edge network at equilibrium theoretically depends on the area elasticity $k_A$, we investigate the influence of $k_A$ on the curvature-induced rigidity. We do so by combining our mean-field model and simulations of the spherical VM.

$k_A$ affects tissue rigidity both 1) locally, by contributing to the T1's energy barrier $\Delta U$, and 2) globally, by altering the equilibrium configurations of the edge network. Locally, Eq.\ref{delta_per} shows that $k_A$ contributes to $\Delta U$ as a proportionality factor in the quartic dependence on the edge length $L$. Since the coefficients $A,B,C \sim \mathcal{O}(1)$ and $k_A \sin^2 \left( \frac{4}{6} \pi \right) \sim \mathcal{O}(10^{-3})$, this contribution is orders of magnitude smaller than the other terms in Eq.\ref{delta_per}. Globally, the effect of $k_A$ can be derived starting from Eqs.~\ref{L-dep}-\ref{tau-dep}. The dependence on $N$ of Eqs.~\ref{spher_cap} and \ref{Z_average}, and hence Eqs.~\ref{L-dep}-\ref{tau-dep}, is based solely on geometric arguments. Therefore, we can assume that the effect of the area elasticity on $\langle L \rangle$ and $\langle \tau \rangle$ (and hence $\langle \tau_S \rangle$) is implicitly incorporated in the term $P_{\text{flat}}(k_A,p_0)$. This result implies that the dependence of the spherical edge length and tension networks on $k_A$ is determined by the dependence of the corresponding flat cell perimeters on $k_A$. It has been previously shown that the area elasticity does not affect the tension percolation threshold $p_{0_{\text{flat}}}^*$ of the flat VM\cite{li2019mechanical}. As such, we expect the distributions of $(L,\tau,\tau_S)$, and hence the curvature-induced shift of tissue rigidity, to be mostly independent of the value of $k_A$.

\begin{figure}
\includegraphics[width=\linewidth]{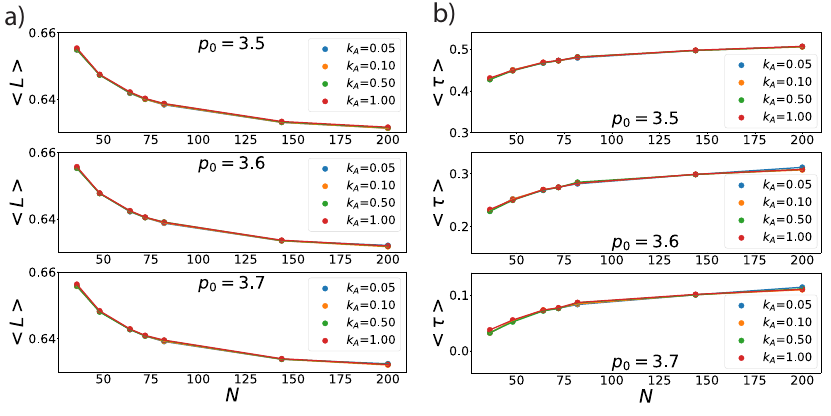}
\caption{a) Average edge length $\langle L \rangle$ and b) tension $\langle \tau \rangle$ versus $N$ at various $p_0= \{ 3.5, 3.6, 3.7 \}$. Different colors correspond to different values of the area elasticity $k_A=\{0.05, 0.1, 0.5,1 \}$.}
\label{new_supp_fig:5}
\end{figure}

\begin{figure}
\includegraphics[width=\linewidth]{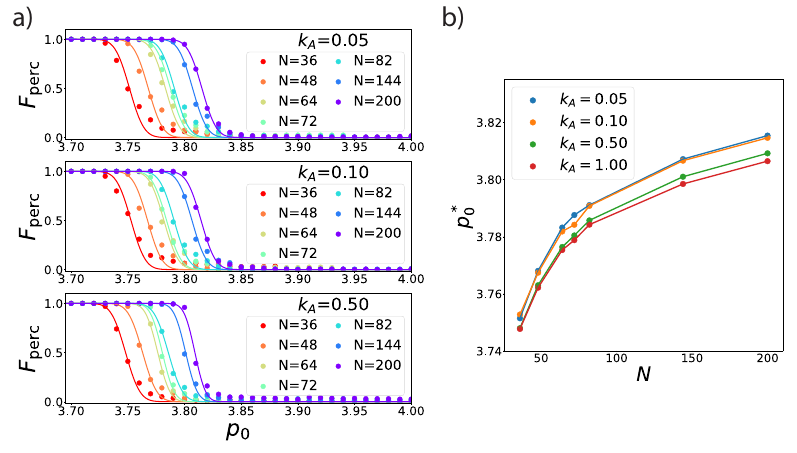}
\caption{a) Probability of tension percolation $F_{\text{perc}}$ versus $p_0$ for different $N$ and $k_A$. $F_{\text{perc}}(p_0,N,k_A)$ is well described by the error function $(1/2)\{1-\text{erf}[(p_0 - \mu_N)/(\sqrt{2}\sigma_N)]\}$ (solid lines). b) Percolation threshold $p_0^*(N)$ versus $N$ for different values of the area elasticity parameter $k_A$. For each $k_A$, $p_0^*(N)$, and hence the onset of rigidity, increases with increasing the number of cells $N$.}
\label{new_supp_fig:6}
\end{figure}

To validate our predictions, we perform additional simulations of the spherical VM with $N=\{36,48,64,72,82,144,200\}$ for $k_A= \{0.05, 0.1, 0.5\}$ and $p_0$ in the range of $3.49-4$. Supp. Fig.\ref{new_supp_fig:5} shows that, at fixed $p_0$, changes in the area elasticity $k_A$ do not impact edge lengths and tensions on average.
Following the same procedure as for $k_A=1$, we compute the tension percolation probability $F_{\text{perc}} (N,p_0,k_A)$ and the percolation threshold  $p_0^*(N,k_A)$. Supp. Fig.\ref{new_supp_fig:6}a shows that $F_{\text{perc}}(N,p_0,k_A)$ recapitulates the behavior observed at $k_A=1$. $F_{\text{perc}} (N,p_0,k_A )$ is well-described by the error function and shifts toward lower $p_0$ as $N$ decreases, regardless of the chosen $k_A$. Remarkably, $p_0^*(N,k_A)$ exhibits a similar dependence on $N$ across different values of the area elasticity (Supp. Fig.\ref{new_supp_fig:6}b) and the impact of $k_A$ is modest ($\max ( \Delta p_0^* ) = 0.009$). This supports our hypothesis that, to the leading order, the role of $k_A$ is negligible. It is worth noting that increasing $k_A$ slightly decreases the rigidity transition. This increase likely reflects effects of the area elasticity not accounted for in our first-order approximation. Further studies are necessary to determine in more detail the influence of area elasticity on curved epithelia.

Similar to $k_A$, cellular density might also affect the plasticity of our spherical layer. We explore these effects by simulating the spherical VM at different cell packing densities $\rho$. 
\begin{figure*}
\includegraphics[width=\linewidth]{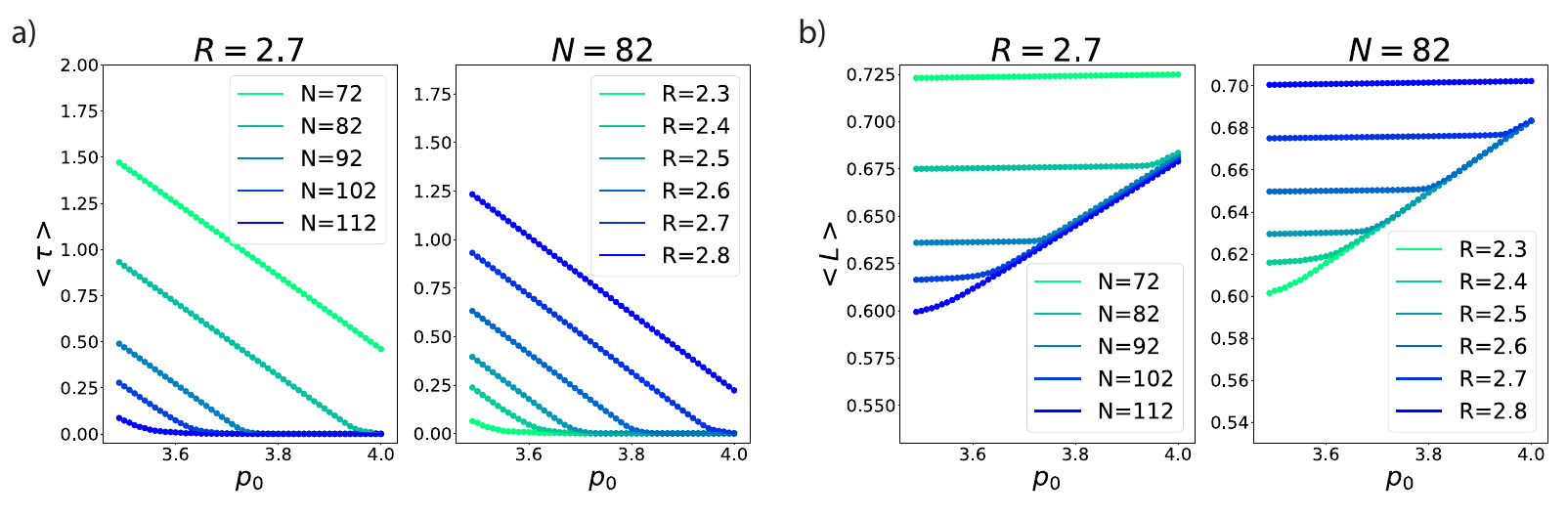}
\caption{a) Average edge tension $\langle \tau \rangle$ and b) average edge length $\langle L \rangle$ versus $p_0$ in spherical VMs with various cell densities ($k_A=1$). (Left panels) Different colors represent spheres with different $N$ and fixed radius $R=2.7$, corresponding to cell densities $\rho$ ranging $0.786-1.22$. (Right panels) Different colors represent spheres with different $R$ and fixed $N=82$, corresponding to cell densities $\rho$ ranging $0.83-1.233$.}
\label{new_supp_fig:7}
\end{figure*}
We vary $\rho$ by fixing the sphere radius $R$ while changing the number of cells $N$ and vice versa. Supp. Fig.\ref{new_supp_fig:7}a shows the average tension $\langle \tau \rangle$ as a function of $p_0$ for fixed $R=2.70$ and $N=\{72,82,92,102,112\}$, corresponding to cell densities in the range $\rho=0.786-1.22$, and for fixed $N=82$ and $R=\{2.30,2.40,2.50,2.60,2.70,2.80\}$, corresponding to cell densities in the range $\rho=0.83-1.233$. $k_A$ is set at $k_A=1$. Similar to the flat VM, higher cell densities coincide with lower edge tensions on average. $\langle \tau \rangle$ vanishes at lower $p_0$ values with increasing $\rho$, indicating a shift in the onset of rigidity. This is also evident from the plot of the average edge length versus $p_0$ (Supp. Fig.\ref{new_supp_fig:7}b). In spherical VMs with higher densities, $\langle L \rangle$ deviates from the constant plateau of the rigid phase at lower $p_0$.

\begin{figure}
\includegraphics[width=\linewidth]{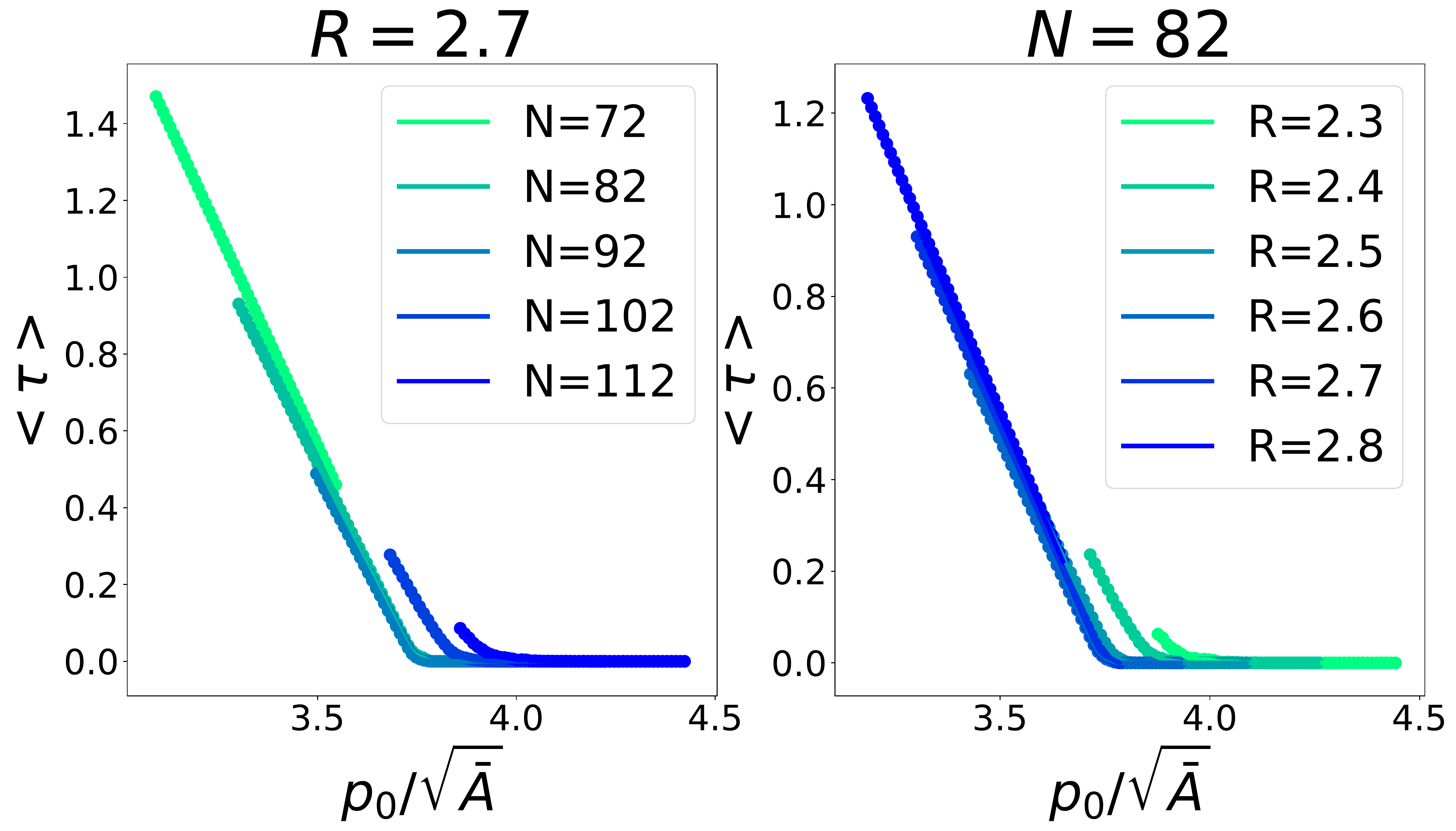}
\caption{Average edge tension $\langle \tau \rangle$ versus the rescaled target shape index $p_0/\sqrt{\bar{A}}$, with $\bar{A}$ the average cell area ($k_A=1$). (Left panel) Different colors represent spherical VMs with different $N$ and fixed sphere radius $R=2.7$. $\langle \tau \rangle$ collapses onto a single curve for $N \mdm{\leq} 92$ (coinciding with $\rho \sim 1$). (Right panel) Different colors represent spherical VMs with different $R$ and fixed number of cells $N=82$. $\langle \tau \rangle$ collapses onto a single curve for $R \mdm{\geq} 2.5$ (coinciding with $\rho \sim 1$).}
\label{tau_collapse_rho}
\end{figure}

We ask whether this density-dependent shift of edge tensions is trivially due to the change in the average cell area $\bar{A}$. Interestingly, after rescaling the target shape index $p_0$ by $\sqrt{\bar{A}}=\sqrt{4\pi R^2 /N}$, $\langle \tau \rangle$ collapses onto a single curve only for densities $\rho < 1$ (Supp. Fig.\ref{tau_collapse_rho}).
This collapse is consistent regardless of the protocol used to vary the cell density. Conversely, when $\rho>1$, tissue rigidity is affected by additional non-trivial factors beyond the change in the average cell area.

\section{Dynamic vertex model of cell monolayers on a spherical surface}
In a recent work by some of us\cite{tang2022collective}, we have adapted our previously published dynamic vertex model (DVM) on flat surfaces\cite{mitchel2020primary} to study cellular dynamics in a confluent spherical monolayer. Here we utilize this spherical DVM to probe the effect of substrate curvature on the motility-driven unjamming transition. The model and the calculated dynamical quantities are described briefly below.

\textit{Energy Hamiltonian and single-cell motility}---
The DVM Hamiltonian is described by the functional form of Eq.\ref{energy}. Cell motility is described by the overdamped equation of motion:
\be
\Gamma \frac{d  \vec{r}_i}{dt}  = \vec{F}_i^{\text{int}} + v_0 \hat{n}_i,
\label{eq:overdamped}
\ee
where $\vec{r}_i(t)$ is the position of the i-th vertex at time $t$ and $\Gamma$ is the frictional damping coefficient. The forces on the vertices arise from two sources: interactions between cells mediated by cell-cell contacts ($\vec{F}_i^{\text{int}}$) and cellular tractions applied on the substrate ($v_0 \hat{n}_i$). The first contribution is determined by the spatial gradient of the tissue mechanical energy defined in Eq.\ref{energy}, whereas the second contribution is modelled as an active motility force applied on vertex $i$ with constant strength $v_0$. The direction of the traction force $\hat{n}_i$ is determined by the sum of polarization vectors of the three cells sharing the vertex. The implementation of individual cell polarization is similar to other self-propelled particle models on spherical geometries\cite{sknepnek2015active}. The polarization vector of the m-th cell is given by $\hat{n}_m=\left( \cos \theta_m,\sin \theta_m \right)$, where the angle of polarization $\theta_m$ also exhibits an over-damped dynamics:
\be
\frac{d \theta_m}{d t}= \eta, \;\;\;\;\;\;\; \langle \eta_i(t)\eta_j(t')\rangle = 2D_r\delta_{ij}\delta(t-t').
\label{eq:theta}
\ee

$\eta$ is a random variable obeying the statistics of Gaussian white noise with zero mean and variance $2D_r$. This description recapitulates the front-back polarity of migrating cells\cite{bi2016motility,szabo2010collective,ladoux2016front}. $D_r$ is the rate at which the orientation of the cell polarization vector randomizes. At any point $\vec{x}$ on the sphere and at every time step, the displacements and the polarizations of the vertices positions are projected on the corresponding tangent plane. Given a vector $\vec{y}$ at a point $\vec{x}$ on the sphere, the projection operator $\vec{P}_T (\vec{x},\vec{y})$ is defined as\cite{sknepnek2015active}:
\be
\vec{P}_T (\vec{x},\vec{y}) = \vec{y} - \left (\hat{x} \cdot \vec{y} \right) \hat{x}
\ee
This exercise ensures that cells remain constrained on the spherical surface throughout the entire simulation. 

\textit{Simulation details}---Simulations are initiated from the zero-motility ($v_0=0$) equilibrium configurations obtained through the minimization procedure described above. Given these configurations, finite motility ($v_0>0$) simulations are then performed. The vertices locations are updated through the Euler method according to Eq.\ref{eq:overdamped}. The time step of the simulations is $\Delta t=0.02 \tau$, where $\tau=\Gamma/K_P$ represents the unit of time in the DVM. We use Eq.\ref{eq:theta} to determine the rotational noise on the cell polarizations. 

In our simulations, T1 rearrangements are allowed with an embargo timer of $0.2\tau$ when an edge is shorter than $l_c=0.05\sqrt{A_0}$\cite{das2021controlled}. We use Surface-Evolver\cite{brakke1992surface} to implement the DVM simulations. We perform DVM simulations of spheres with number of cells $N$ ranging $30-250$. Analogous to the spherical VM, the sphere radius $R$ is chosen such that $R=\sqrt{N/4\pi}$. We examine the effect of curvature by varying the motility force $v_0$ in the range $0.02-0.14$ and keeping fixed the target shape index $p_0=3.65$, the rotational noise strength $D_r=1$, and the area elasticity $k_A=1$. In this way, the single-cell dynamics remains Brownian. For each parameter set, we perform 10 independent simulations initiated from different random cell configurations.

From each simulation, we compute two different dynamical quantities. 
\begin{figure}
\includegraphics[width=\linewidth]{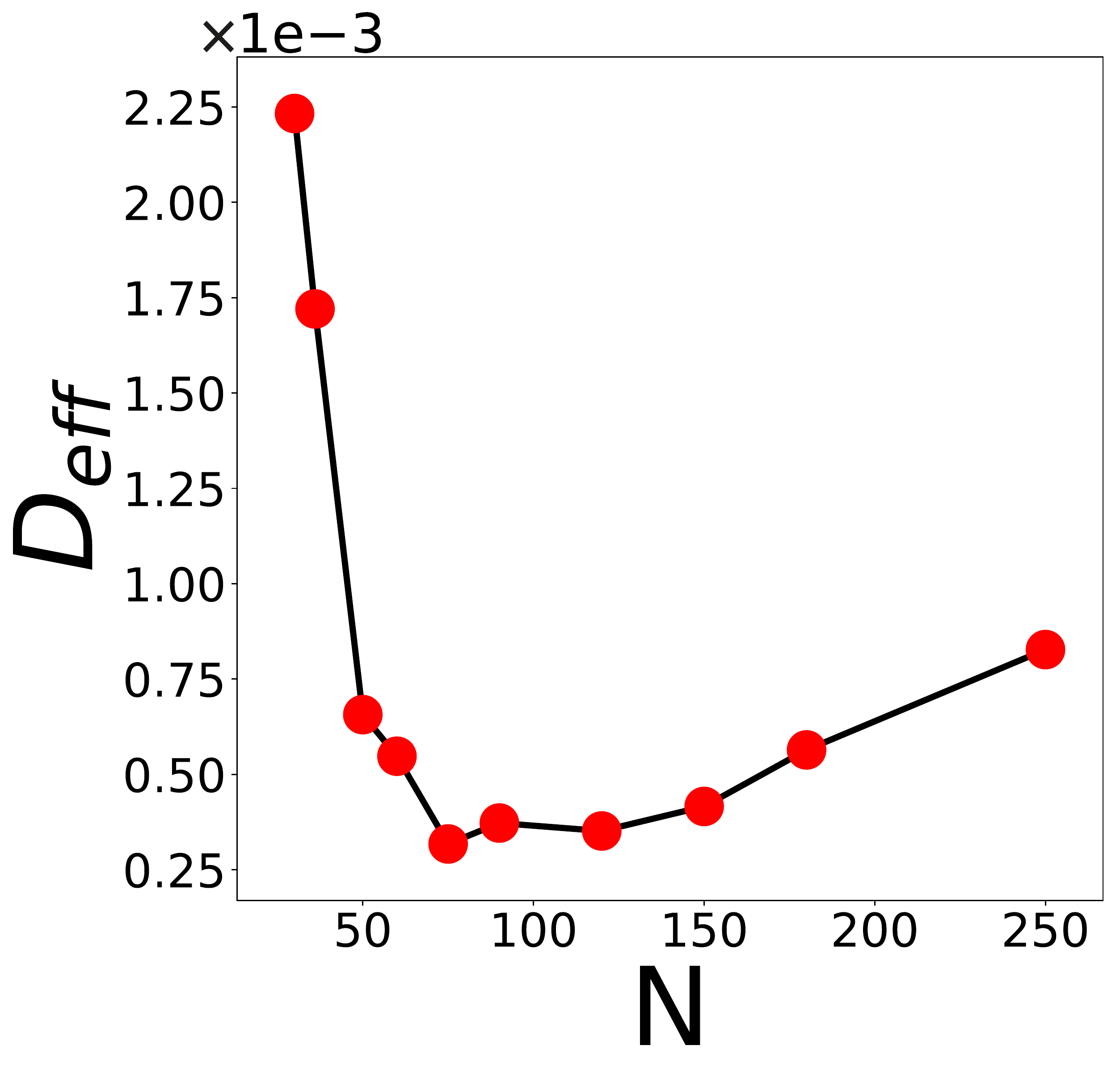}
\caption{Diffusion coefficient $D_{\text{eff}}$ versus $N$ at fixed $v_0=0.09$ ($p_0=3.65$, $D_r=1$, $k_A=1$).}
\label{supp_fig:5}
\end{figure}
First, the average mean-square displacement (MSD) $\langle \Delta r^2(t)\rangle$ of cells as a function of time. In the MSD, distance is calculated based on the length of the geodesic curve connecting the initial and final position of the cell in the time interval $t$. We quantify the long-time diffusive behavior of the MSD by extracting the diffusion coefficient $D_{\text{eff}}$. We calculate $D_{\text{eff}}$ as the ratio of the MSD to the time lag at the largest time lag. \mdm{$D_{\text{eff}}$ provides a dimensionless, normalized measure of particle diffusivity that allows for direct comparisons across systems with different parameters.} Supp. Fig.\ref{supp_fig:5} shows the dependence of $D_{\text{eff}}$ on the radius of curvature at $v_0=0.09$. $D_{\text{eff}}$ exhibits a strong dependence on $N$, decreasing dramatically with increasing $N$ in the range $30-150$. A slight increase is observed for $N>150$ without exceeding the threshold of $10^{-3}$ that marks the fluid-to-solid transition\cite{sahu2020small,bi2016motility}. As shown in Fig.\ref{fig:5}b, we estimate $D_{\text{eff}}$ for different values of $v_0$ and $N$. We also estimate the T1 transition's rate $k_{T1}$, defined as the number of neighbor exchanges per junction per unit of time.
\end{document}